\documentclass[twocolumn,amsmath,amssymb,floatfix,10pt,aip,jcp]{revtex4-1}

\newcommand{\bvec}[1]{{\mathbf{\string#1} }}
\newcommand{\upd}{\mathrm{d}}
\usepackage{graphicx}
\usepackage{amssymb,amsfonts,amsmath}
\usepackage{color}
\usepackage{ulem}
\usepackage{fancybox}

\newcommand{\rr}{\boldsymbol{r}}
\newcommand{\nab}{\boldsymbol{\nabla}}

\newcommand{\taua}{\tau_\text{a}}

\newcommand{\ttau}{\tilde{\tau}}
\newcommand{\Da}{\mathcal{D}_\text{a}}

\DeclareSymbolFont{matha}{OML}{txmi}{m}{it}
\DeclareMathSymbol{\varv}{\mathord}{matha}{118}
\newcommand{\Vext}{\varv}

\newcommand{\Hess}{\nab\nab}

\newcommand{\p}[1]{p^\text{(#1)}}
\newcommand{\pact}[1]{p_\mathrm{act}^\text{(#1)}}
\newcommand{\pactb}{p_\mathrm{act}}

\begin{document}

\title{Pressure, surface tension and curvature in active systems: A touch of equilibrium}

\author{Ren\'e Wittmann}
\email{rene.wittmann@hhu.de}
 \affiliation{Department of Physics, University of Fribourg, CH-1700 Fribourg, Switzerland}
 \affiliation{Institut f{\"u}r Theoretische Physik II, Weiche Materie,
Heinrich-Heine-Universit{\"a}t D{\"u}sseldorf, D-40225 D{\"u}sseldorf, Germany}
\author{Frank Smallenburg}
 \affiliation{Institut f{\"u}r Theoretische Physik II, Weiche Materie,
Heinrich-Heine-Universit{\"a}t D{\"u}sseldorf, D-40225 D{\"u}sseldorf, Germany}
 \affiliation{Laboratoire de Physique des Solides, CNRS, Univ. Paris-Sud, Univ. Paris-Saclay, Orsay, France.}
\author{Joseph M.\ Brader}
 \affiliation{Department of Physics, University of Fribourg, CH-1700 Fribourg, Switzerland}

\date{\today}

\begin{abstract}
We explore the pressure of active particles on curved surfaces and its relation to other interfacial properties.
We use both direct simulations of the active systems as well as simulations of an equilibrium system with effective (pair) interactions designed to capture the effects of activity. 
Comparing the active and effective passive systems in terms of their bulk pressure, we elaborate that the most useful theoretical route to this quantity is via the density profile at a flat wall.
This is corroborated by extending the study to curved surfaces
and establishing a connection to the particle adsorption and integrated surface excess pressure (surface tension).
In the ideal-gas limit, the effect of curvature on the mechanical properties can be calculated analytically in the passive system with effective interactions, and shows good (but not exact) agreement with simulations of the active models.
It turns out that even the linear correction to the pressure is model specific and equals the planar adsorption in each case, which means that a known equilibrium sum rule can be extended to a regime at small but nonzero activity.
In turn, the relation between the planar adsorption and the surface tension is reminiscent of the Gibbs adsorption theorem at an effective temperature.
At finite densities, where particle interactions play a role, the presented effective-potential approximation captures the effect of density on the dependence of the pressure on curvature.
\end{abstract}

\maketitle

\section{Introduction}

Recent advances in colloidal science have allowed for the creation of active colloids:
synthetic particles which are capable of using energy from their environment to fuel active, self-propelled motion \cite{palacci2013living, buttinoni2013,  theurkauff2012dynamic}.
Due to their constant motion, systems of active particles are inherently out-of-equilibrium, and hence do not follow the usual rules of equilibrium thermodynamics. 
The emergence of activity has spurred a new interest in the statistical physics of such systems \cite{takatori2015, cates_tailleur2014}.
A topic of particular interest is the question whether equilibrium concepts, such as pressure \cite{takatori2014, winkler2015, solon2015EOS}, interfacial or surface tension \cite{speck_interface2014,paliwal2017,Das2019},
chemical potential \cite{paliwal2018chemical,stenhammar2013, meer2016chempot}, temperature \cite{loi2008effective,szamel2014}, and free energy \cite{solon2018generalized,stenhammar2013,wittmannbrader2016}, 
can be extended to provide meaningful insights in active systems as well. Of these quantities, the active pressure has perhaps been scrutinized the most. 
In its most simple definition, the pressure can be identified with the force per unit area the active particles exert on the confining walls. Unlike in equilibrium,
this force generally depends not only on the bulk properties, but also on the wall-particle interaction \cite{solon2015EOS}, preventing the definition of a bulk pressure in this way.

The standard model system of active Brownian particles (ABPs) consists of particles which propel themselves with a constant velocity along their instantaneous orientation, subject to rotational Brownian motion.
Such ABPs interact with each other and the wall only via isotropic interactions.
In this special case, the pressure has been shown to be a state function, which provides one condition to predict coexistences between different phases, analogous to the equilibrium pressure~\cite{solon_BrownianPressure2015}.
This means that the pressure that a fluid of ABPs exerts on a flat wall is simply equal to the bulk pressure, regardless of the wall-particle interaction. 
However, if the wall is curved, it is not obvious how the force per unit surface area is related to the active bulk pressure.
Indeed, simulations have shown that, even for a noninteracting active gas, the active pressure of ABPs 
strongly depends on the wall curvature \cite{smallenburg2015} and follows the structure of the wall in a local fashion \cite{nikola2016}.
This observation can be explained by considering regions of a highly negative (positive) curvature as cavities (obstacles), 
which lead to an increased (decreased) probability of finding a particle near the wall and thus a higher (lower) contribution to the wall pressure compared to its bulk value. 
Here and throughout this paper we use the convention that the surface normal points towards the bulk.

Neglecting all autocorrelation functions of the self-propulsion force beyond second order, 
the activity of ABPs can be approximately represented by colored noise \cite{faragebrader2015}.
These active Ornstein-Uhlenbeck processes (AOUPs) constitute another example for an isotropically interacting model system, in which the self-propulsion vector fluctuates in both direction and length. In a manner of speaking, AOUPs are even more simplistic and closer to equilibrium \cite{fodor2016} than ABPs, as their motion reverts to simple overdamped Brownian dynamics in the limit of short correlation time (without additionally introducing a Brownian thermal noise).
Although there are, in general, some important differences between the two model systems~\cite{szamel2014,solonEPJST,fily2017,Das2018},
many essential aspects of the nonequilibrium behavior of ABPs and AOUPs are quite similar.  For example, recent results concerning the pressure of AOUPs at curved walls \cite{sandford2017,sandford2018} qualitatively reproduce the observations for ABPs \cite{nikola2016}.

The AOUPs model is also known for being a convenient starting point to develop
an effective description of active systems by means of their configurational probability distribution,
allowing to exploit techniques familiar for (near) equilibrium systems~\cite{maggi2015sr,marconi2015,marconi2016,faragebrader2015,sharma2017,SpeckCRIT}.
As a possible second step to this equilibrium mapping, the effective-potential approximation (EPA)~\cite{activePAIR,faragebrader2015,marconi2016mp,activePRESSURE,wittmannbrader2016,SpeckCRIT} has been employed to construct a closed theory to study the active system, e.g., using variational methods. 
The crucial idea of this approximation is to derive a pairwise-additive effective interaction force to represent the activity within a framework developed for passive systems.
This procedure is even possible if the two particles considered have different activities~\cite{activeMixture}.

The basic idea of representing active particles by equilibrium ones has an ambiguous taste.
On the one hand, the simplicity of the time-evolution equation allows for the construction of
particularly simple theories, on the other hand, many inherently out-of-equilibrium aspects cannot be accounted for in this way.
Nevertheless, this approach was proven to be quite useful in several situations,
since some steady-state results can be accurately reproduced in systems with low activity and spatial dimensionality~\cite{activePAIR,marconi2016mp}.
In the small-activity limit, the effective equilibrium mapping recovers several exact results for an ideal gas \cite{marconiExactpressure2017,activeMixture}.
For interacting particles, some closed formulas for the mechanical properties have been derived~\cite{marconi2016,activePRESSURE}, which are consistent with the concepts of swim pressure \cite{takatori2014} and active interfacial tension~\cite{speck_interface2014}.
In addition, the EPA provides a solid qualitative understanding of the phase behavior of interacting active systems~\cite{faragebrader2015,wittmannbrader2016}.
 Recently, the effective equilibrium reasoning has been adopted for other models \cite{paoluzzi2018}
and some alternative approaches have been proposed to obtain improved one-body distribution functions \cite{caprini2018,fily2018}.
A quantitative description of active particles in effective equilibrium is, however, usually difficult,
in particular when it comes to a calculation involving the pair correlations in an interacting three-dimensional system~\cite{SpeckCRIT}. A related argumentation in a different context expounds that the predictions of an approximate theory become worse if the results are obtained via two-point instead of one-point distributions \cite{archerevans2017}.

Another important question related to the applicability of the EPA 
concerns the role of curvature, which emerges in two distinct types.
First, the notion of a \textit{potential curvature} describes the change of 
slope of a soft potential landscape, i.e., the change of magnitude of the external force, in a certain direction. 
It has been concluded in the context of various one-dimensional problems that
most accurate results can be obtained for a small absolute value of the potential curvature~\cite{sharma2017,caprini2018,fily2018}. 
Second, in higher spatial dimensions, the shape of a hard wall or particle can be characterized by its \textit{geometrical curvature}.
More generally, one can also refer to a characteristic equipotential line when the interaction is soft.
The first proper prediction of the qualitative dependence on geometrical curvature is
that a larger number of active ideal particles accumulate in a cavity than at an obstacle \cite{marconi2015}. Later, an explicit analytic result has been obtained for the density of particles trapped in a cavity~\cite{fily2017}. In this case, the theory has been confirmed to become exact in the limit of an infinite persistence time, which has been reported to be generally the case in one dimension \cite{faetti1988,UCNA}. At an obstacle or for interacting particles, where the geometrical curvature is positive, the EPA cannot be employed properly without an empirical correction \cite{activePAIR}.

In this paper we address the issues outlined above specific to the EPA~\cite{maggi2015sr, marconi2015, marconi2016, faragebrader2015, wittmannbrader2016, sharma2017, activePAIR, activePRESSURE, fily2017, marconi2016mp}
in more detail and in the context of a well-studied property of active particles, namely their pressure.
Explicitly, we are concerned with the fundamental questions:
(i) how accurately can we predict the active pressure in the presence of interparticle interactions,
(ii) to what degree can the peculiar behavior of active particles at curved \cite{smallenburg2015,sandford2017,sandford2018} or structured \cite{nikola2016,sandford2018} surfaces be captured, 
(iii) does the (corrected) theory also provide proper results for a positive geometrical curvature, and 
(iv) what is the relation between the pressure and the surface excess properties at the wall? 
We corroborate our theoretical findings by performing computer simulations of active systems.
To allow for a quantitative comparison, we go beyond approximate theories to implement the EPA by performing explicit simulations of the effective passive system.
We conduct our study in two dimensions, since the accuracy of the effective potentials is known to decrease with increasing dimensionality \cite{activePAIR}.
Moreover, in two dimensions, overdamped models of active particles, which ignore hydrodynamic interactions, are more realistic, since a substrate can act as a momentum sink.

The remainder of this paper is arranged as follows:
In Sec.~\ref{sec_theory} we briefly recapitulate the effective equilibrium model and the EPA~\cite{activePAIR} to the extent required here.
We then compare in Sec.~\ref{sec_bulk} active and passive simulations to measure the pressure of an interacting system in the bulk and on a flat wall.
In Sec.~\ref{sec_curved}, we consider active ideal gases near curved walls, and describe at small activity (or curvature) the relations between active pressure, (excess) adsorption and surface tension
(or, more accurately, the \textit{negative integrated surface excess pressure})\cite{clarifyST}, reminiscent of equilibrium sum rules \cite{sumrules}. Moreover, we extract the leading-order curvature correction to the pressure in the presence of interactions.
We conclude in Sec.~\ref{sec_conclusions} on the perspective of the employed equilibrium mapping and the 
relation between 
 bulk and surface excess properties
in active systems with isotropic interactions.

\section{Effective equilibrium theory \label{sec_theory}}

In the following we briefly introduce the main results of
the EPA required to later calculate the mechanical pressure.
Throughout the paper we assume that in the active system there is no translational thermal noise, which would be necessarily present in a passive Brownian system,
and are only interested in the steady-state behavior.
Then both schemes, based on the one-dimensional Fox approach~\cite{FOX} and Unified Colored Noise Approximation~\cite{UCNA},
to develop a generalized equilibrium mapping for the multicomponent system in an arbitrary dimension are equivalent \cite{activePAIR}.
The reader interested in the full derivation and further technical details
is referred to the extensive literature on this subject, in particular, Refs.~\onlinecite{faragebrader2015,maggi2015sr,activePAIR}.
The related microscopic equations of motion of ABPs and AOUPs are explained in Appendix~\ref{app_sim}.

In effective equilibrium, we consider an active system whose steady-state configurational probability distribution $P_N(\bvec{r}^N)$ 
solves the equation \cite{activePAIR} 
\begin{equation}
\beta\bvec{F}_i P_N-\sum_j^N\nab_j(\mathcal{D}_{ji}\color{black}P_N)=0\,,
\label{eq_FP}
\end{equation}
 where $\beta=(k_\text{B}T)^{-1}$ is the inverse of the temperature $T$ with Boltzmann's constant $k_\text{B}$.
Note that in the absence of thermal noise, $\beta$ here simply functions as an (activity-independent) inverse energy unit, whose choice does not affect the behavior of the system.
Moreover, $\bvec{F}_i(\rr^N)$ represents the conservative force on particle $i$ and the
 dimensionless effective diffusion tensor $\mathcal{D}_{ij}(\rr^N)$
serves to represent the activity of the $N$ particles of diameter $d$
and equals unity in the passive case.
 Explicitly, it depends on the persistence time $\taua$ of the self-propelled motion and its magnitude. 
The latter is characterized by the active diffusivity $D_\text{a}$ (in the case of AOUPs)
or by the constant self-propulsion velocity $v_0$ (for ABPs),
where, in two dimensions, we can identify both parameters according to the relation $D_\text{a}= v_0^2\taua/2$.
For later convenience, we also introduce the persistence length $l_\text{p}:=\sqrt{2D_\text{a}\taua}=\taua v_0$ of the active motion.

Explicitly, the components of the inverse of $\mathcal{D}_{ij}$ read \cite{activePAIR}
\begin{equation}
 \mathcal{D}_{ij}^{-1}(\rr^N)=\Da^{-1}
 \left(\boldsymbol{1}\delta_{ij}-\tau d^2
 \nab_i 
\beta\bvec{F}_j(\rr^N)\right)
 \label{eq_Deff}
\end{equation}
with the dimensionless persistence time $\tau=\taua/\tau_0$, where $\tau_0 = (\beta\gamma d^2)$ denotes the damping time, 
and diffusivity $\mathcal{D}_\text{a}=D_\text{a}\beta\gamma$, where $\gamma$ is the friction coefficient.  
Inspecting Eq.~\eqref{eq_Deff}, we see that 
 in the absence of external forces active particles experience an effective active temperature scale $\beta_\text{eff}=\beta/\Da$ since their diffusion is enhanced by $\Da$ \cite{szamel2014}
compared to a passive system, which is explicitly recovered in the white-noise limit of the AOUPs model, $\tau\rightarrow 0$ while $\Da=1$ is kept finite.
This effective
 diffusion is reduced by approaching a repulsive wall or when particles with repulsive interactions accumulate.
While this intuition already reflects the behavior of active particles quite nicely in a dynamical picture,
the versatility of the effective equilibrium approach comes from the possibility to describe the nonequilibrium steady states
by means of a static formula, i.e., Eq.~\eqref{eq_FP}, which still depends on this (effective) diffusion tensor.
As detailed later, its contribution results in an increase of effective attraction or a decrease of active pressure when the particles become more active.
As a further consequence, the effective dynamics of active particles are described by a complex interplay of both the activity-dependent effective diffusion
and modified force terms, which determine the effective equilibrium state.
 In fact, depending on the chosen theoretical framework, a slightly different interpretation of the latter is necessary to specify a Fokker-Planck equation for the time evolution of $P_N$, where $\mathcal{D}_{ij}$ acts as a diffusion tensor:
 the Fox approximation suggests the existence of an effective force \cite{activePRESSURE},
 while the Unified Colored Noise Approximation results in an additional contribution to the bare interaction force \cite{marconi2015}.
 The explicit form of Eq.~\eqref{eq_Deff} is the same in both theories, as long as translational Brownian noise is negligible.
  Hence, the steady-state condition, Eq.~\eqref{eq_FP}, is identical and the two different interpretations of the force terms are formally equivalent, so that we may choose the most convenient one \cite{activePAIR}.

Returning to the static behavior, 
we solve as a first step Eq.~\eqref{eq_FP} for the effective equilibrium probability distribution $P_N$.
 Then we can readily identify effective interaction potentials (see appendix~\ref{app_pots}) accounting for the 
increase of probability to find a repulsive active particle near a boundary or another particle.
Considering the case with $N=2$ particles, we have $\bvec{F}_1=-\bvec{F}_2=-\nab u(r)$ with the pair potential $u(r)$.
Then we can define the effective pair potential $\beta u^\text{eff}(r)=-\ln P_2$.
Likewise, from the interaction force $\bvec{F}_1=-\nab \Vext(\bvec{r})$ of a single particle ($N=1$) with an external one-body field $\Vext(\bvec{r})$,
we obtain $\beta \Vext^\text{eff}(\bvec{r})=-\ln P_1$.
 Notice that the effective diffusion tensor in Eq.~\eqref{eq_Deff} is not always positive definite.
As detailed in appendix~\ref{app_curv}, it may become negative for potentials with a negative potential curvature or
a positive geometrical curvature.
Hence, to be able to extend our study to the behavior of active particles at obstacles in this work, we
employ in appendix~\ref{app_pots} an empirical modification, the \textit{inverse-$\tau$ approximation},  
which ensures qualitatively correct behavior of effective potentials \cite{activePAIR} even in such situations.
For a cavity, the effective diffusion tensor is always positive definite.

In order to make analytic progress, we follow Ref.~\onlinecite{fily2017} and choose a simple power-law dependence of the bare interaction potentials of the form $\sim\!\!\lambda x^n$, introduced in full detail in appendix~\ref{app_pots}, with an integer-valued exponent $n\geq2$ and a softness parameter $\lambda$, which ranges between 0 (no interaction) and infinity (hard interaction).
The most handy potential, one branch of a parabola, results in a spurious discontinuity of the effective potentials at the position of the vertex~\cite{fily2017},
since the second derivative of a parabola does not vanish at the apex.
Despite this artifact, it can be verified by choosing exponents $n>2$ that the analytic results at a hard wall obtained in this way remain invariant.
In order to avoid any pitfalls all numerical calculations are carried out with the exponent $n=4$.

\section{Active bulk pressure \label{sec_bulk}}

The pressure $\pact{B}$ in a torque-free active system can be measured in bulk \cite{winkler2015},
or from the force on a flat wall in a sufficiently large system, which we denote as $\pact{W}\equiv\pact{B}$. 
At the moment, the usefulness of the EPA to calculate the active pressure is not quite evident.
This is mostly due to the misjudgment that the desired quantity can be identified with the \textit{effective} thermodynamic pressure $p_\text{eff}$ 
 obtained from a standard equilibrium calculation for a passive system interacting with the effective potential $\beta u^\text{eff}(r)$.
For example, using the virial theorem, we have
\begin{align}
\beta p_\text{eff}&= \rho_0 -\frac{\pi}{2}\rho_0^2\!\int_0^\infty\upd r\, r^2\,g(r)\,\frac{\partial\beta u_\text{eff}(r)}{\partial r}\,, \label{eq_pVeff}  
\end{align}
where $g(r)$ is the radial distribution function and $\rho_0$ the bulk density.
However, this effective pressure was explicitly shown \textit{not} to share obvious attributes of a (mechanical) active pressure~\cite{SpeckCRIT,activePRESSURE}.
The reasons for this discrepancy have been discussed in Ref.~\onlinecite{activePRESSURE}
and an artificial rescaling was presented on a formal level
(this rescaled pressure $\p{R}$ was argued to be inferior to the virial pressure $\p{V}$ introduced below).
On the other hand, the EPA provides a convenient theoretical route to access the radial distribution function of the active particles,
which can be used as input for a closed virial-like expression to calculate the active pressure.
Moreover, we will propose another, more intuitive way to calculate the pressure within the EPA
by its force exerted on a planar (and later curved) wall.

Using the virial theorem, statistical
formulas for the active pressure
(and interfacial or surface tension) have been derived in Refs.~\onlinecite{activePRESSURE,marconi2016}, which depend solely on properties of the bulk fluid via the ensemble average $\boldsymbol{\mathcal{D}}(\bvec{r})$ of the effective diffusion tensor $\mathcal{D}_{ij}$.
We use the approximate representation 
\begin{align}
\mathcal{D}_\text{a}\boldsymbol{\mathcal{D}}^{-1}(\bvec{r})\approx
\boldsymbol{1}+\tau d^2 \int\upd\bvec{r}'\frac{\rho^{(2)}(\bvec{r},\bvec{r}')}{\rho(\bvec{r})}\Hess \beta u(\bvec{r},\bvec{r}')
\label{eq_Dav}
\end{align}
of this quantity, where $\rho^{(2)}$ is the two-particle density, which in the bulk becomes $\rho^{(2)}(|\bvec{r}-\bvec{r}'|)\simeq\rho_0^2g(r)$.
The choice of the expression in Eq.~\eqref{eq_Dav} can be motivated in two ways.
The first strategy involves an expansion up to linear order in the persistence time (low-activity approximation) to be able to carry out the ensemble average of $\mathcal{D}_{ij}$~\cite{marconi2016,activePAIR}
and replace this average with Eq.~\eqref{eq_Dav} to restore in the resulting expressions the neglected higher-order terms.
The second approximation amounts to rederive the virial formulas in a more indirect way, which allows to explicitly take the average of the inverse diffusion tensor, i.e., Eq.~\eqref{eq_Deff}~\cite{activePAIR,activePRESSURE}.

To apply the virial theorem to the equality in Eq.~\eqref{eq_FP},
we separate the force in an external part representing the boundary and an internal force due to particle interactions.
Then the virial pressure of an active bulk system follows in two dimensions as~\cite{activePRESSURE,marconi2016}
\begin{align}
\beta \p{V} &=\frac{\mbox{Tr}[{\boldsymbol{\mathcal{D}}}]}{2}\rho_0-\frac{\pi}{2}\rho_0^2\!\int_0^\infty\upd r\, r^2\,g(r)\,\frac{\partial\beta u(r)}{\partial r}\,. \label{eq_pV}  
\end{align}
The second term equals the passive virial, compare Eq.~\eqref{eq_pVeff}, and only depends implicitly on the activity through changes in the (effective) radial distribution $g(r)$ compared to a passive system. 
The trace in the first term can be written as
\begin{align}
\mbox{Tr}[{\boldsymbol{\mathcal{D}}}]&= \frac{2 \mathcal{D}_\text{a}}{1 + \pi \rho_0 \tau d^2 \!\int\upd r\, r g(r)\left(\frac{\partial^2\beta u(r)}{\partial r^2}+ \frac{1}{r}\frac{\partial\beta u(r)}{\partial r}\right) }\cr
&= \frac{2 \mathcal{D}_\text{a}}{1 + \bigg\langle \frac{\tau d^2}{N}\sum\limits_{i<j}  \left(\frac{\partial^2\beta u(r_{ij})}{\partial r_{ij}}+ \frac{1}{r_{ij}}\frac{\partial\beta u(r_{ij})}{\partial r_{ij}}\right)\!\bigg\rangle } \label{eq_pV2}
\end{align}
since we consider a homogeneous and isotropic system.
The derivation of Eq.~\eqref{eq_pV} circumvents the definition of effective interaction potentials.
Therefore, the expression for $\p{V}$ can also be used together with the radial distribution obtained from computer simulations of a true active system.
In this case, we write $\pact{V}$, whereas $\p{V}$ corresponds to a calculation within the EPA.
Note that the bulk pressure $\pact{B}$ is also obtained from a virial-based approach \cite{winkler2015}
but should not be confused with the approximate expression for $\pact{V}$
and that it is not possible to determine an expression in the EPA that is analog to $\pact{B}$.

Apart from the bulk route we now consider an active fluid at a planar wall characterized by the bare external potential $\Vext(x)$.
For such a setup we can deduce the mechanical pressure
\begin{align}
\beta \p{W}=-\int_{-\infty}^\infty\upd x\,\frac{\partial\beta\Vext(x)}{\partial x}\,\rho(x)
\label{eq_pW}
\end{align}
from its most fundamental definition: the force per unit area exerted on a wall.
Again, the inhomogeneous one-body density $\rho(x)$ can readily be 
measured for an active system, yielding $\pact{W}$, or for a passive system within the EPA, where we write $\p{W}$. 
In the latter case, it is important to determine $\rho(x)$ for the effective wall with $\Vext^\text{eff}(x)$, although the pressure is then measured with the help of $\Vext(x)$.
For isotropically interacting  active particles, which we aim to describe here, $\pact{W}$ is independent of the wall potential~\cite{solon_BrownianPressure2015,nikola2016}, and hence equal to $\pact{B}$. Therefore, we set $\pactb \equiv \pact{W}\equiv\pact{B}$.
For an active ideal gas, all expressions
\begin{align}
 \beta \p{W}=\beta \p{V}=\beta \pact{W}=\beta \pact{V}=\beta \pact{B}=\Da\rho_0
 \label{eq_pid}
\end{align}
yield the exact ideal swim pressure~\cite{activePRESSURE}.
Apparently from Eq.~\eqref{eq_pVeff}, the effective thermodynamic pressure $\beta p_\text{eff}\!=\!\rho_0$, on the other hand,
is independent of both activity and the wall-particle interaction.

In general, there exists no trivial relation between $\p{W}$ and $p_\text{eff}$. 
Inspired by the equivalence in Eq.~\eqref{eq_pid} for an ideal gas, it is instructive to
 multiply the effective pressure with $\Da$, i.e., switching to the effective temperature scale $\beta_\text{eff}$. 
To make the connection with Eq.~\eqref{eq_pW} we replace $\Vext(x)$ with the effective external potential $\Vext^\text{eff}(x)$    
and define the effective-temperature pressure
\begin{align}
\beta \p{T}= -\Da\int_{-\infty}^\infty\upd x\,\frac{\partial{\beta\Vext^\text{eff}(x)}}{\partial x}\,{\rho(x)}\equiv \beta_\text{eff}\, p_\text{eff}
\label{eq_pT}
\end{align}
for a flat wall and a sufficiently large system.
The latter equality follows from the wall theorem of equilibrium thermodynamics, 
which holds for any interacting passive fluid if the calculation can be done exactly.
Alternatively, $p_\text{eff}$ can equally be determined via Eq.~\eqref{eq_pVeff}.
By construction, the correct (active) ideal-gas solution $\beta p^\text{(T)}=\Da\rho_0$ 
is also recovered from Eq.~\eqref{eq_pT}. 
Since this definition explicitly makes use of an EPA result ($\Vext^\text{eff}$ or $p_\text{eff}$), 
there is no sensible equivalent for the full active system.

\begin{figure}  \centering
\includegraphics[width=0.5\textwidth] {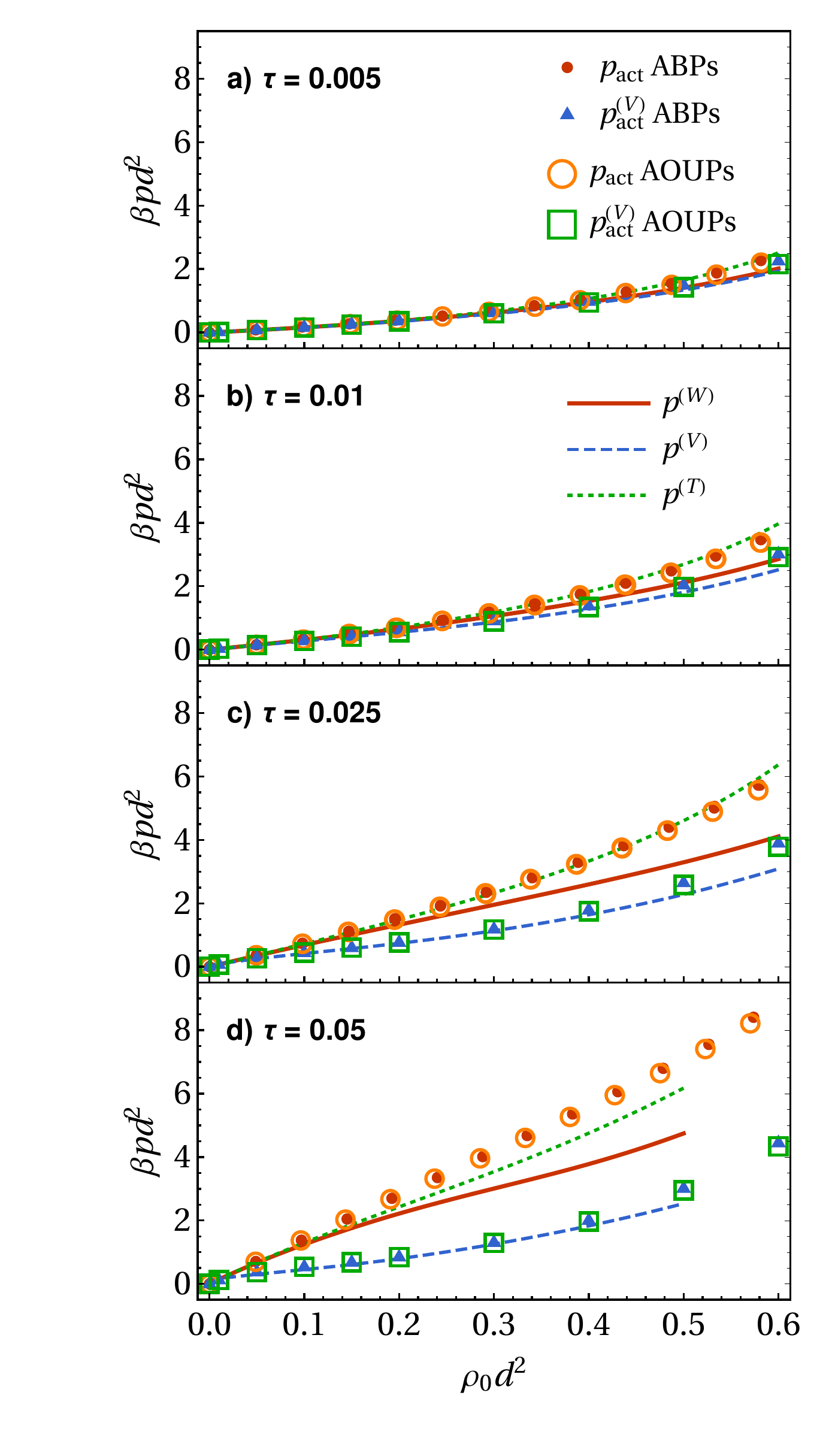}
\vspace{-1cm}
\caption{
Comparison between different methods for determining the active pressure of bulk systems, as a function of the bulk density $\rho_0$ and for different rotational diffusion times
{\bf (a)}~$\tau \!=\! 0.005$, {\bf (b)}~$\tau \!=\! 0.01$, \textbf{c)}~$\tau \!=\! 0.025$ and {\bf (d)}~$\tau \!=\! 0.05$.
In all cases, the self-propulsion speed is fixed at $v_0 = 24  d / \tau_0$. 
The points represent measurements performed directly in simulations of ABPs or AOUPs, and lines indicate the EPA pressures measured in a passive system with effective pair interactions. 
The label $p_\text{act}$ collects the equivalent reference results for $\pact{B}$ and $\pact{W}$. 
For $\tau \!=\! 0.05$, the passive system phase separates at densities $\rho_0 d^2 \gtrsim 0.5$.
}
\label{fig_p0}
\end{figure}

\subsection{Interacting particles at a flat wall \label{sec_planar}}

In order to better assess the accuracy of the EPA we now extend the comparison from Ref.~\onlinecite{activePRESSURE} of the different routes to calculate the active pressure by
(i) implementing the effective pair potentials numerically in a passive simulation to circumvent the need for further approximations,
(ii) additionally considering the expressions $\p{W}$ and $\p{T}$ proposed in Eq.~\eqref{eq_pW} and Eq.~\eqref{eq_pT}, respectively, for the pressure from the force of a wall
and (iii) including different active computer simulation results as a reference,
where we (iv) also test the general value of Eq.~\eqref{eq_pV} for an active system by calculating $\pact{V}$.
 Also recall that in the present study we focus on two-dimensional systems.
 In all simulations we fix the self-propulsion speed of the particles $v_0 = 24  d / \tau_0$, while changing their persistence time $\tau$ and hence the associated persistence length $l_\text{p}$.
 An increase of $\tau$ thus represents an increase of activity.

In Fig.~\ref{fig_p0} we show the active pressure as a function of the density for different activity parameters, and compare it to pressures measured in the corresponding passive system with effective interactions. For ABPs and AOUPs, we measured both the bulk pressure $\pact{B}$ in a system without walls (using the virial expression in Ref.~\onlinecite{winkler2015}), and the mechanical pressure $\pact{W}$ on a flat wall in a sufficiently large system. 
As expected, $\pact{B} = \pact{W}$ in all cases, representing the true pressure $\pactb$ exerted by the active particles on their container. Moreover, we find essentially the same active pressures for the ABPs and AOUPs models for all investigated activities and densities.
In contrast, the pressure $\pact{V}$ derived from the effective diffusion tensor is approximate, and increasingly deviates from the true bulk pressure as activity increases, for both the ABPs and AOUPs models.

\begin{figure} [t] \centering
\includegraphics[width=0.475\textwidth] {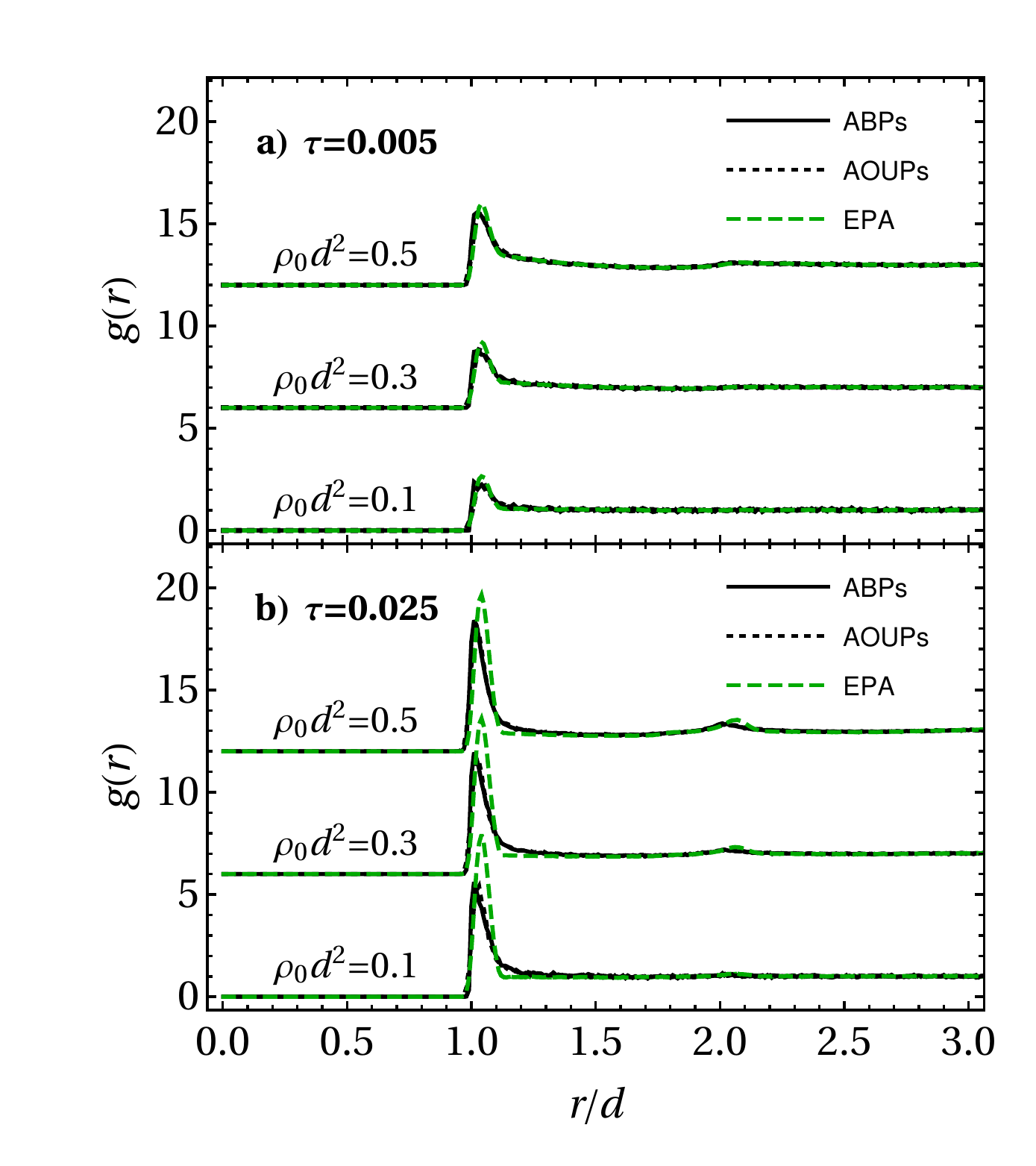}\vspace*{-0.2cm}
\caption{
Comparison of the radial distribution function $g(r)$ in the passive and active system, for different persistence densities $\rho_0$ as indicated, at fixed self-propulsion velocity $v_0 = 24  d / \tau_0$. 
The persistence times are \textbf{(a)}~$\tau \!=\! 0.005$ and {\bf (b)}~$\tau \!=\! 0.025$.}
\label{fig_gr}
\end{figure}

For the corresponding passive systems, we also plot in Fig.~\ref{fig_p0} the pressures $\p{V}$, $\p{W}$, 
and $\p{T}$ using Eqs.~\eqref{eq_pV}, \eqref{eq_pW}, and \eqref{eq_pT}, respectively.
The theoretical pressures $\p{W}$, calculated from the bare potential $\Vext(x)$, and $\pact{V}$, 
calculated from the effective diffusion tensor, exhibit a similar behavior at low activity $\tau \lesssim 0.01$,
whereas the (rescaled) thermodynamic pressure $\p{T}$ of the passive system is always larger.
 At higher activities, there are significant differences between the three theoretical methods and, quite
surprisingly, $\p{T}$ follows the true pressure $\pactb$
much more closely, even at higher densities.
Significant deviations only occur for strong activity $\tau \gtrsim 0.05$, 
where the passive system
undergoes a phase separation for densities $\rho_0 d^2 \gtrsim 0.5$ and hence pressures can only be reliably calculated up to that density. 
This phase transition shifts to lower densities when further increasing $\tau$~\cite{faragebrader2015,wittmannbrader2016}.
Also the agreement between $\p{W}$ and $\pactb$ remains reasonable at low densities or small activities.

Comparing the results of the virial pressures $\p{V}$ of the passive system (calculated using the bare interparticle potential) 
and $\pact{V}$ of the active system, we find good agreement in all cases where phase separation does not occur. 
Since these are both calculated from the radial distribution function $g(r)$, this observation suggests that the approximations involved in deriving Eq.~\eqref{eq_pV} are cruder than
those leading to the approximate radial distribution $g(r)$ within the EPA.
To check this, we compare $g(r)$ for different parameters in Fig.~\ref{fig_gr},
which illustrates the known deviations at higher densities and activities, although the agreement remains reasonable at all parameters considered.  
Such a comparison has already been done in three dimensions and with another approximation for the effective potential,
where the disagreement was shown to be much more severe~\cite{faragebrader2015,SpeckCRIT}, whereas in one dimension no further approximation becomes necessary and an even better match between theory and simulations was found \cite{marconi2016mp}. 
We again find virtually identical results for $g(r)$ in the ABPs and AOUPs models, and good agreement with the EPA model.

Finally, we observe in Fig.~\ref{fig_p0} a small horizontal offset between the points corresponding to the active pressure $\pactb$ of AOUPs and ABPs simulated at the same particle number and volume, especially at high activity. 
These systems were simulated in the presence of two flat walls, and hence this shift results from a difference in the observed bulk densities, caused by a difference in the adsorption at the wall between these two models. This is intriguing, as the bulk pressures and radial distribution functions of the two models are essentially the same for all densities and activities. Evidently, while ABPs and AOUPs behave identical in the bulk, they show significant differences in their behavior near a wall. 
This observation will be quantified and extended in Sec.~\ref{sec_curved}, where we consider more general systems with curved walls, 
for which the flat-wall results are recovered in the zero-curvature limit.

\section{Curvature dependence \label{sec_curved}}

Having verified that the active pressure in the EPA is best calculated by the force exerted on a wall,
we still need to answer the question whether the effective-temperature pressure $\p{T}$, defined in Eq.~\eqref{eq_pT}, 
is superior to the more realistic mechanical pressure $\p{W}$ from Eq.~\eqref{eq_pW} also in more general situations.
The logical next step is thus to consider curved surfaces, focusing on a circular geometry of radius $R$ for the moment.
To distinguish a cavity (particles inside the circle) from an obstacle (particles outside the circle),
we have to consider 
two different potentials $\Vext^{-}(r)$ and $\Vext^{+}(r)$.
These expressions, as well as the corresponding effective potentials $\Vext^\text{eff$\mp$}(r)$, formally become equivalent in the limit $R\rightarrow\infty$ of a planar wall, see appendix~\ref{app_pots}.
Following the conventzion of Ref.~\onlinecite{smallenburg2015}, we formally consider a signed curvature radius $R$, which becomes negative for a cavity,
to represent the corresponding wall by a negative geometrical curvature, see appendix~\ref{app_curv} for more details.
We denote the respective pressures $\p{W-}(R^{-1})$ for a cavity with $R<0$
and $\p{W+}(R^{-1})$ for an obstacle with $R>0$ by a modified superscript.
With these adjustments, the overall pressure $\p{W}(R^{-1})$ is a continuous function 
where the planar limit $\p{W}(0)=\p{W}$ is given by Eq.~\eqref{eq_pW}.
The same applies to all other quantities considered.

\subsection{Pressure, adsorption and surface tension \label{sec_pgs}}

Calculating the total force on the area (circumference) $A=|2\pi R|$
of a circular wall of radius $R$ with the convention described above, the two contributions to the pressure $\p{W}(R^{-1})$ become
\begin{align}
\beta \p{W$\mp$}(|R^{-1}|)=\pm\int_0^\infty\upd r\,\frac{r}{R}\frac{\partial\beta\Vext^{\mp}(r)}{\partial r}\,\rho(r)\,,
\label{eq_pWofR}
\end{align}
and equally for $\pact{W}(R^{-1})$ if $\rho(r)$ is measured in the active systems.
The argument $|R^{-1}|$ serves to emphasize that we evaluate the right-hand side for the absolute value of $R$, which appears in the potentials specified in appendix~\ref{app_pots}.
To obtain the correct result as a function of the signed curvature $R^{-1}$, we later change the sign of $R$ in the formulas with superscript $(-)$ for a cavity.
The corresponding expressions for 
\begin{align}
\beta \p{T$\mp$}(|R^{-1}|)=\pm\Da\int_0^\infty\upd r\,\frac{r}{R}\frac{\partial\beta\Vext^\text{eff$\mp$}(r)}{\partial r}\,\rho(r)\,,
\label{eq_pTofR}
\end{align}
in the passive system simply follow from replacing $\Vext(r)$ with $\Da\Vext^\text{eff}(r)$ in Eq.~\eqref{eq_pWofR}.

It is instructive to further consider some (mechanical) excess properties at the surface.
Equivalent to the surface excess grand potential in statistical mechanics for a passive system \cite{sumrules},
we define mechanically the total surface tension~\cite{clarifyST} (or , more accurately, the negative integrated excess pressure) of a fluid at a circular surface as
\begin{align}
 \sigma^{(\mp)}(|R^{-1}|)=\int_0^\infty\upd r\,\frac{r}{R}\, (p\,\Theta(\mp(r-R))-\boldsymbol{p}_\text{T}(r))\,.
 \label{eq_gamma}
\end{align}
where we locate the surface at the apex ($r=R$) of the wall potential,
$p$ denotes the bulk pressure and $\boldsymbol{p}_\text{T}(r)$ denotes the component of the pressure tensor tangential to the interface.
Moreover, we 
define the (excess) adsorption 
\begin{align}
\Gamma^{(\mp)}(|R^{-1}|)=\int_0^\infty\upd r\,\frac{r}{R}\,(\rho(r)-\rho_0\, \Theta(\mp(r-R)))\,
 \label{eq_gammaEFF}
\end{align}
from the density profile perpendicular to the wall alone.
The general curvature dependence of this quantity in a hard cavity has already been studied using the EPA in Ref.~\onlinecite{fily2017}.

Within the EPA, the density profile of an active ideal gas is given explicitly by the simple expression $\rho(\bvec{r})\!=\!\rho_0\,\exp(-\beta\Vext^\text{eff}(\bvec{r}))$ in any geometry,
where $\rho_0$ is the bulk density.
The tangential pressure $\boldsymbol{p}_\text{T}(r)=\rho(r)\boldsymbol{\mathcal{D}}_\text{T}(r)$ can be expressed \cite{marconi2015} in terms of the Eigenvalue of $\boldsymbol{\mathcal{D}}$ from Eq.~\eqref{eq_Dav} along the direction tangential to the surface and the bulk pressure of an active ideal gas uniquely follows from Eq.~\eqref{eq_pid}.
With the help of simple power-law potentials specified in appendix~\ref{app_pots} we can
easily study the influence of the softness of the interaction specified by the parameter $\lambda$ (see Appendix~\ref{app_pots}) entering as a prefactor
and evaluate the hard-wall limit, $\lambda\rightarrow\infty$, to derive simple analytic results.
To do so for an obstacle, we formally replace the lower boundary of the radial integrals with minus infinity.

 At this point, let us remind ourselves of certain sum rules \cite{sumrules} which provide a relation between the quantities defined above in equilibrium, which we denote by the subscript ``eq''.
The adsorption follows from the surface tension via the Gibbs adsorption theorem
\begin{align}
  \Gamma_\text{eq}=-\frac{\partial\sigma_\text{eq}}{\partial\mu_\text{eq}}\stackrel{\text{id}}{=}-\beta\sigma_\text{eq}
   \label{eq_Gibbsadsorption}
\end{align}
where $\mu_\text{eq}$ is the chemical potential and the last equality, providing an explicit $\mu_\text{eq}$-independent relation, holds for an ideal gas.
Moreover, the curvature dependence of the pressure
\begin{align}
p_\text{eq}(R^{-1})=p_\text{eq}+\frac{\sigma_\text{eq}}{R}+\frac{\partial\sigma_\text{eq}}{\partial R} 
\label{eq_itsumrule}
\end{align}
  follows in equilibrium from the bulk pressure and curvature-dependent surface tension $\sigma_\text{eq}(R^{-1})$.
 We stress that, like in Eq.~\eqref{eq_Gibbsadsorption}, the derivation of this relation
requires the notion of a well-defined chemical potential \cite{sumrules}, which does not exist in a nonequilibrium active system.

 Returning to the EPA for an active system, we can simply introduce via Eq.~\eqref{eq_Gibbsadsorption} an
effective surface tension 
 \begin{align}
   \beta\sigma^\text{eff}(R^{-1})=-\Gamma(R^{-1})
   \label{eq_sigmaEFF}
\end{align}
 for an ideal gas in a radially symmetric (effective) external field.
While the adsorption from Eq.~\eqref{eq_gammaEFF} can also be used as a quantifier for the active system, 
such an effective surface tension is apparently different from $\sigma$ in Eq.~\eqref{eq_gamma},
 which contains an additional factor $\boldsymbol{\mathcal{D}}_\text{T}$.
Note that for an interacting system both theoretical expressions $\sigma$ and $\sigma^\text{eff}$ contain additional terms \cite{activePRESSURE}.
 In general, with the help of Eq.~\eqref{eq_itsumrule}, the effective surface tension $\sigma^\text{eff}(R^{-1})$ 
can be further used to calculate a curvature-dependent generalization of the effective pressure $p_\text{eff}$.
Rescaling the resulting function $p_\text{eff}(R^{-1})$ with the effective temperature, we define from the surface-tension route the effective-temperature pressure
\begin{align}
 \p{$\tilde{\text{T}}$}(R^{-1}):=\Da p_\text{eff}(R^{-1})\,,
 \label{eq_pTofRsigma}
\end{align}
which is not the same quantity as $\p{T}(R^{-1})$ from Eq.~\eqref{eq_pTofR}, obtained from the force route.
This is because $\Vext^\text{eff}(r)$ is not a function of ($r-R$) alone. Note that
 $\p{$\tilde{\text{T}}$}(R^{-1})$ can also be obtained by replacing the derivative $\partial/\partial r$ with $-\partial/\partial R$ in Eq.~\eqref{eq_pTofR}.

In our active simulations, we used a different route to calculate the total surface tension in the planar case, which only relies on the globally averaged pressure. This avoids the explicit numerical integration of the pressure profile over the simulation box. We verified that this approach yields the same results as Eq.~\eqref{eq_gamma} in the planar limit.
Introducing a global pressure tensor $\boldsymbol{P}$ which formally contains an explicit normal contribution due to the surrounding walls, we define
\begin{align}
 \sigma_\text{act}(0)= \frac{L_y}{2}(\boldsymbol{P}_\text{N}-\boldsymbol{P}_\text{T})\,, 
 \label{eq_gamma0}
\end{align}
  through the anisotropy of the global pressure tensor in a system confined between two walls (hence the factor two) parallel to the $x$-axis of the box, separated by a distance $L_y$. Here, we define the pressure tensor $\boldsymbol{P}$ as
 \begin{equation}
  \boldsymbol{P} = \boldsymbol{P}_\mathrm{swim} + \boldsymbol{P}_\mathrm{vir} + \boldsymbol{P}_\mathrm{wall}, 
\end{equation}  
where the three terms on the right-hand side represent contributions from the swim pressure, the pair interactions, and the wall interaction, respectively. For the first term, we generalize the expressions proposed by Winkler {\it et al.} to tensor form. This results in 
\begin{eqnarray}
  \boldsymbol{P}_\mathrm{swim} =   \frac{1}{V D_r} \left\langle \sum_{i=1}^N   \mathbf{f}^\mathrm{tot}_{i} \mathbf{v}^\mathrm{act}_i \right\rangle, \label{eq:actpressswim}
\end{eqnarray}
where $\mathbf{f}^\mathrm{tot}_{i}$ represents the total force on particle $i$, including the self-propulsion force, and $\mathbf{v}^\mathrm{act}_i$ denotes the self-propulsion part of its velocity.
The pair interaction term is given by
\begin{eqnarray}
\boldsymbol{P}_\mathrm{vir} = \frac{1}{V} \left\langle \sum_{<i,j>}   \mathbf{f}_{ij} \mathbf{r}_{ij} \right\rangle. \label{eq:actpressvir}
\end{eqnarray}
 Finally, the wall term is obtained by treating the walls as two additional particles of infinite mass \cite{varnik2000molecular}, located at $y$-positions $y_{w,1}$ and $y_{w,2}$. This yields
\begin{eqnarray}
  \boldsymbol{P}_\mathrm{wall} =  -\frac{\mathbf{e}_y \mathbf{e}_y}{V} \left\langle \sum_{i=1}^N \sum_{j=1}^{2}  \nu'(y_i - y_{w,j}) \times (y_i - y_{w,j}) \right\rangle\,.\ \ \  \label{eq:actpresswall}
\end{eqnarray}
Note that for an active ideal gas, $\boldsymbol{P}_\mathrm{vir}$ vanishes.

\subsection{Active ideal gas at a hard wall}

If a fluid is in contact with a curved wall, its surface properties may depend on its curvature.
This is already the case for a passive ideal gas, if the interaction potential differs from that of a hard wall.
Introducing activity, we even expect a (nonlocal \cite{nikola2016,sandford2018}) curvature dependence in the hard-wall limit \cite{smallenburg2015}.

\subsubsection{Uniformly curved walls\label{sec_circular}}

Let us first consider an active ideal gas in a hard circular cavity and at a hard circular obstacle of radius $R$.
The following analytic predictions of the theory are obtained by calculating the density profile of the corresponding passive systems and taking the hard-wall limit of the expressions defined in Sec.~\ref{sec_pgs}.  
Similar results for a spherical wall in three dimensions are discussed in appendix~\ref{app_circular}.
For our two-dimensional system, we find the explicit formulas
\begin{align}
\frac{p^\text{(W-)}(R^{-1})}{ \Da\rho_0}&=1-\frac{\sqrt{\pi}}{2}\frac{l_\text{p}}{R}\,,\cr
\frac{p^\text{(W+)}(R^{-1})}{ \Da\rho_0}&=1-\frac{\sqrt{\pi}}{2}\frac{l_\text{p}}{R}+\frac{l_\text{p}^2}{R^2}+\mathcal{O}\left(\frac{l_\text{p}^3}{R^3}\right),
\label{eq_cavobs}
\end{align}
 for the pressure from Eq.~\eqref{eq_pWofR} and
\begin{align}
\frac{\Gamma^{(-)}(R^{-1})}{\rho_0}&=\frac{\sqrt{\pi}}{2}l_\text{p}-\frac{1}{2}\frac{l_\text{p}^2}{R}\,,\cr
\frac{\Gamma^{(+)}(R^{-1})}{\rho_0}&=\frac{\sqrt{\pi}}{2}l_\text{p}-\frac{1}{2}\frac{l_\text{p}^2}{R}+\frac{\sqrt{\pi}}{4}\frac{l_\text{p}^3}{R^2}+\mathcal{O}\!\left(\frac{l_\text{p}^4}{R^3}\right)\ \ \ \ 
\label{eq_cavobsGAMMA}
\end{align} 
for the adsorption from Eq.~\eqref{eq_gammaEFF}.
All expressions depend only on the persistence length $l_\text{p}$, i.e., they are
independent of the particular choices of persistence time and self-propulsion velocity,
as expected from computer simulations of ABPs \cite{smallenburg2015}.
We could not obtain a full analytic solution for an obstacle; the results stated above follow from a
Taylor expansion in $R^{-1}$ before integrating over the normal coordinate and taking the hard-wall limit.
A numeric evaluation of the wall pressure is easily possible without a noticeable error and we will refer to this case as a nearly hard wall.

\begin{figure} [t] \centering
\includegraphics[width=0.45\textwidth] {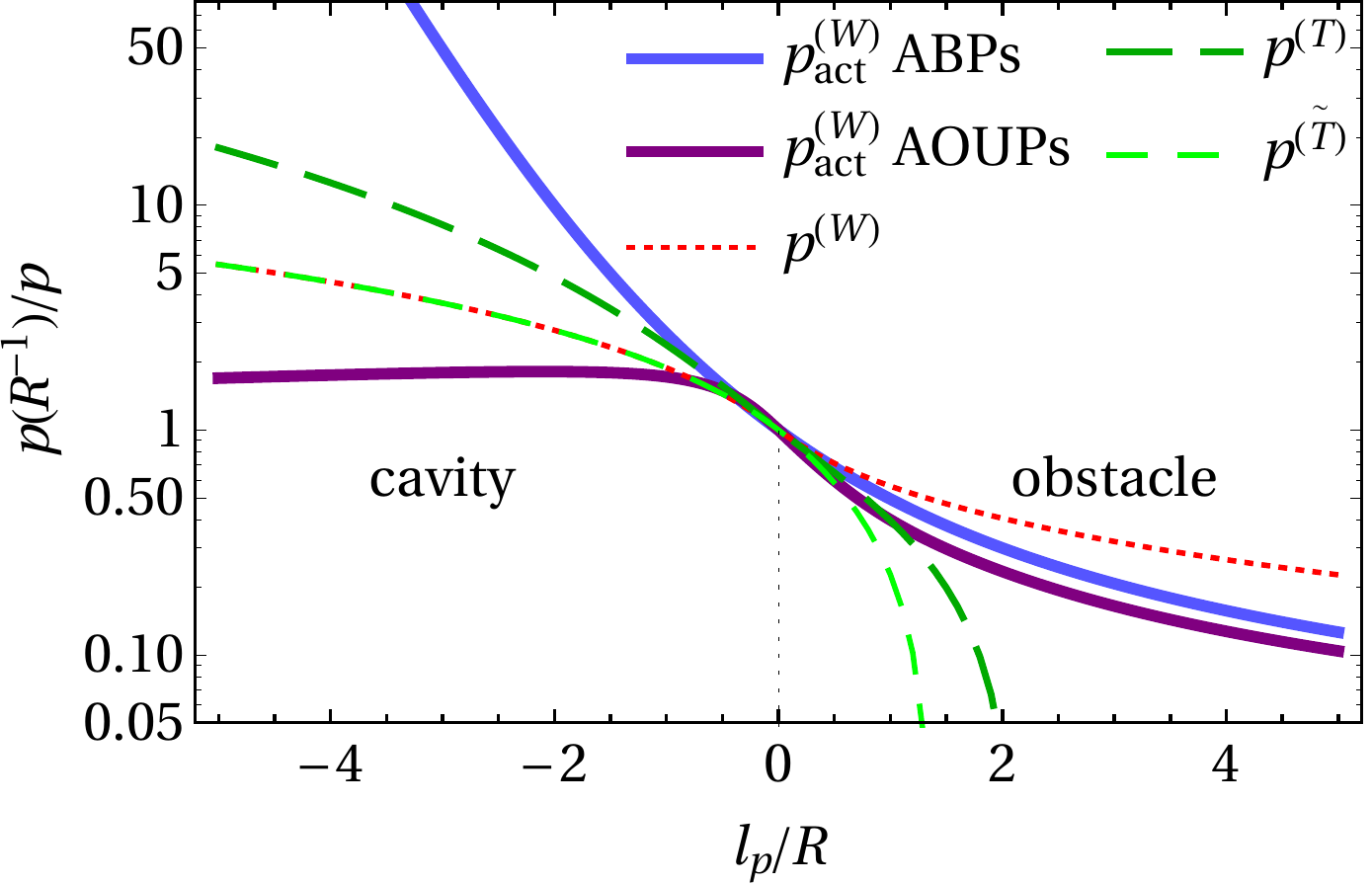}  
\caption{
Pressure $p(R^{-1})$ of an active ideal gas on a (nearly) hard circular wall as a function of the ratio of persistence length $l_\text{p}$ and signed curvature radius $R$.
 We normalize by the pressure $p$ in the corresponding bulk, sufficiently far away from the wall.
All theoretical and numerical curves collapse onto the same line independent of the particular values of the activity parameters.
We compare the active simulation results $\pact{W}$ for ABPs (thick blue line) and AOUPs (thick violet line)
to the EPA results $\p{W}$, calculated according to Eq.~\eqref{eq_pWofR}.
We also include the effective results $\p{T}$ and $\p{$\tilde{\text{T}}$}$ from Eq.~\eqref{eq_pTofR} and Eq.~\eqref{eq_pTofRsigma}, respectively,
which diverge to $-\infty$ for $R>0$.
}\label{fig_idC}
\end{figure}

In a cavity, the expressions for both pressure in Eq.~\eqref{eq_cavobs} and adsorption in Eq.~\eqref{eq_cavobsGAMMA} terminate after the term linear in the inverse radius of the cavity, i.e. its curvature,
whereas at an obstacle we find higher-order terms in the expansion.
The constant and linear term are, however, equivalent in each case.
Therefore, the theoretical pressure $\p{W}(R^{-1})$ and adsorption $\Gamma(R^{-1})$ are smooth functions of $R^{-1}$,
i.e., we find the same slope at $R^{-1}=0$ when approaching the planar-wall limit with an infinite curvature radius from either side.
Unlike the effective surface tension $\beta\sigma^\text{eff}(R^{-1})=-\Gamma(R^{-1})$, the active surface tension 
\begin{align}
 \frac{\beta\sigma}{\rho_0}=-\Da\frac{\sqrt{\pi}}{2}l_\text{p}
 \label{eq_cavobsSIGMA}
\end{align}
obtained from Eq.~\eqref{eq_gamma} is independent of the curvature
but equals $\sigma^\text{eff}(0)$ for a flat wall (up to the factor $\Da$, indicating the different temperature scale).
The first two terms in the expansions for $\p{T}(R^{-1})$ and $\p{$\tilde{\text{T}}$}(R^{-1})$ are the same as in Eq.~\eqref{eq_cavobs}, but the higher-order terms are different, which we illustrate in the following.

We compare in Fig.~\ref{fig_idC} the different theoretical results for the pressure to active simulations.
At an obstacle, the EPA result of Eq.~\eqref{eq_pWofR} for the pressure measured at the true wall exhibits the expected trend, $\p{W}(\infty)=0$, observed for active particles (both ABPs and AOUPs)
to approach zero in the limit of a very small obstacle (or for highly persistent particles).
The prediction of a positive definite pressure is a quite powerful feature of 
the EPA (including the inverse-$\tau$ approximation).
This becomes apparent when regarding the rescaled effective results of Eq.~\eqref{eq_pTofR} or Eq.~\eqref{eq_pTofRsigma}, measured at the effective wall,
which are negative for large values of $R^{-1}$.
 Although this clearly does not match the behavior of the pressure of the active systems, it is understandable how this negative pressure arises in the passive approximation (where the activity only enters through rescaling).
Physically, the effective wall pressure $\p{T}(R^{-1})$ or $\p{$\tilde{\text{T}}$}(R^{-1})$ 
represents the force that a passive particle exerts on a curved sticky hard wall. 
As the effective interaction includes an attractive well, growing a sufficiently small obstacle up to a critical size allows more particles to be adsorbed without sacrificing much free volume, resulting in a negative pressure in this regime. 
In contrast, in the real active Brownian case, the pressure on a repulsive obstacle is always positive~\cite{smallenburg2015}, due to the lack of attractive interactions.
In this situation there is a clear difference between an active system and a passive one with attractive interactions,
which underlines that Eq.~\eqref{eq_pWofR} and, therefore Eq.~\eqref{eq_pW}, is the more robust (and consistent) method to calculate the true active pressure in the EPA,
even though the rescaling of the effective pressure appears to give more accurate results in Fig.~\ref{fig_p0} for an interacting system at a flat wall.

In a cavity, the overall situation is a little more complicated, since the results strongly depend on the particular choice of the model.
This is best illustrated by the significantly different results for ABPs and AOUPs in Fig.~\ref{fig_idC} if $R<0$.
Most notably, the ratio of the wall pressure and bulk pressure $\pact{W}(R^{-1})/\pact{W}$ of AOUPs approaches zero for small cavities, whereas for ABPs the ratio diverges, as first described in Ref.~\onlinecite{smallenburg2015}.
The reason for this behavior is the chosen normalization, since the bulk pressure $\pact{W}$ scales linearly with the density $\rho_0$ of particles that remain in the bulk, compare Eq.~\eqref{eq_cavobs}. 
For a cavity of fixed radius containing a fixed number of particles, taking the limit of $l_\text{p} \rightarrow\infty$ results
in a scenario where essentially all particles are trapped at the wall and push outwards, leading to a finite wall pressure. 
However, the bulk fraction has been shown \cite{fily2017} to decrease exponentially with the persistence length for ABPs and by a power law with exponent $-2/3$ for AOUPs.
 In the latter case, this means that the bulk pressure $\Da\rho_0\propto l_\text{p}^{1/3}$ is still divergent for infinitely
persistent particles, and hence $\pact{W}(R^{-1})/\pact{W}$ vanishes in this limit.
With this in mind, it comes as no surprise that also the theory, for which the bulk density follows a power law with exponent $-2$, does not agree with either model.

To ensure a large enough bulk so that the effects discussed above can be neglected,
we will focus in our further analysis on the term linear in the inverse curvature radius,
which we define in general as
\begin{align}
 m^p:=\left.\frac{\partial p(R^{-1})}{p\,\partial R^{-1}}\right|_{R^{-1}=0}\,
 \label{eq_IS}
\end{align}
 with $\beta p=\Da\rho_0$ in the ideal case.
Moreover, with the pressure universally depending on $l_\text{p}/R$, this initial slope
also represents the leading-order correction in activity, which has been of recent interest due to its proximity to equilibrium~\cite{fodor2016} and exactly solvability in some cases~\cite{marconiExactpressure2017,activeMixture}.
The theoretical result $m^p/l_\text{p}=-\sqrt{\pi}/2\approx-0.886$ for an ideal gas is the same with all possible definitions of pressure within the EPA
and agrees reasonably well with the both active results $m_\text{act}^p/l_\text{p}\approx-0.836$ for ABPs and $m_\text{act}^p/l_\text{p}\approx-1.06$ for AOUPs.

\begin{figure} [t] \centering
\includegraphics[width=0.45\textwidth] {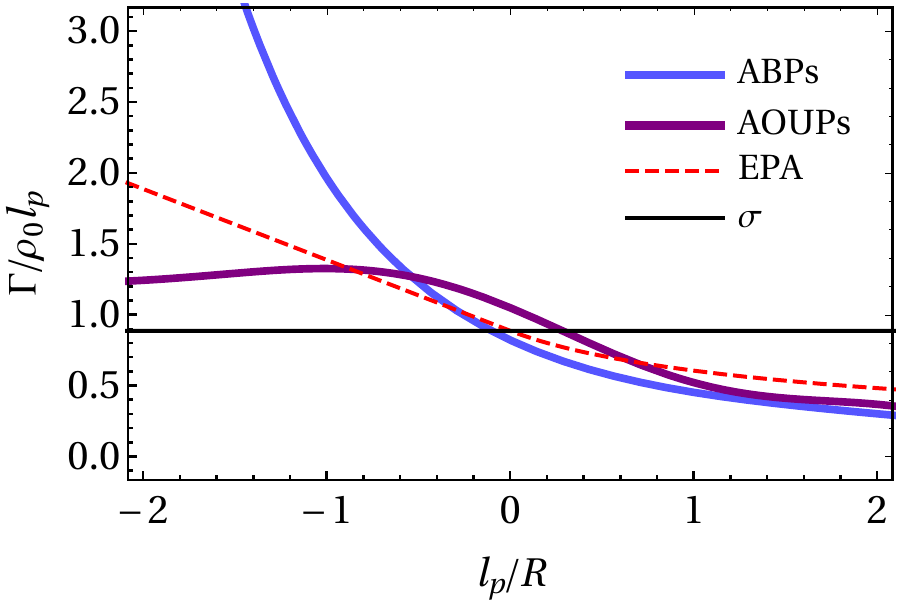}
\caption{
Adsorption $\Gamma$
of an active ideal gas on a circular wall as a function of the ratio of persistence length $l_\text{p}$ and signed curvature radius $R$. Here we use a linear scale to put more emphasis on the nearly-flat-wall behavior.
The horizontal line shows the constant theoretical result for the active surface tension $-\beta\sigma/(\Da\rho_0l_\text{p})$ from Eq.~\eqref{eq_cavobsSIGMA},
which is equal to the adsorption at a planar wall.
}\label{fig_idCgamma}
\end{figure}

To better understand these differences and also the behavior of the active pressure in the normalization of Eqs.~\eqref{eq_cavobs} and~\eqref{eq_IS},
we also analyze the adsorption in Fig.~\ref{fig_idCgamma}.
 For the moment, unlike in Ref.~\onlinecite{fily2017}, we do not normalize this quantity with respect to the particles sitting at the surface.
Using the bulk fraction $\rho_0$ instead, we make a similar observation as for the pressure: the adsorption decreases to zero at small obstacles in all approaches
and there are significant discrepancies between AOUPs and ABPs in a cavity of decreasing size.
For ABPs, the adsorption on the wall diverges exponentially in small cavities, consistent with the exponential depletion of the bulk\cite{fily2017}. In contrast, for AOUPs the dimensionless adsorption $\Gamma/\rho_0 l_\text{p}$ shows a maximum for cavities with a size on the order of the persistence length.

  \begin{table*}[t]
\begin{tabular}[t]{lllll}
\hline\hline
 model &  $m^p/l_\text{p} $ & $-\Gamma(0)/(\rho_0l_\text{p})$ & $\beta\sigma(0)/(\Da\rho_0l_\text{p})$ & $ m^\Gamma/l_\text{p}^2 $ \\
 \hline
   ABPs & $-$0.836$\pm$0.005 \ \ \ \ \ \  & $-$0.836$\pm$0.01 \ \ \ \ \ \  &$-$0.841$\pm$0.01\ \ \ \ \ \ \ &  $-$0.63$\pm$0.05 \\ 
   AOUPs & $-$1.06$\pm$0.03 & $-$1.05$\pm$0.03 & $-$1.02$\pm$0.05 & $-$0.6$\pm$0.1 \\ 
   EPA (numeric) \ \ \ \ & $-$0.886 & $-$0.886 & $-$0.886 & $-$0.5 \\ 
   EPA (exact) & $-\sqrt{\pi}/2$ & $-\sqrt{\pi}/2$ & $-\sqrt{\pi}/2$ & $-$1/2 \\ 
 \hline \hline
     \end{tabular}
  \caption{Comparison of the initial slope $m^p$ of the active pressure, the adsorption up to linear order in $R^{-1}$ (planar adsorption $\Gamma(0)$ and initial slope $m^\Gamma$) 
  and surface tension $\sigma(0)$ of an active ideal gas at a planar hard wall. 
  In both simulations of ABPs and AOUPs as well as the analytic theory based on the EPA, we obtain values which are coherent with the equilibrium sum rules from Eqs.~\eqref{eq_Gibbsadsorption} and~\eqref{eq_itsumrule}
  at leading order in the inverse curvature radius within each respective model.
  }
  \label{tableSUM}
  \end{table*}

Calculating the initial slope 
\begin{align}
 m^\Gamma:=\left.\frac{\partial \Gamma(R^{-1})}{\rho_0\,\partial R^{-1}}\right|_{R^{-1}=0}
 \label{eq_ISgamma}
\end{align}
of the adsorption
we find the value $m_\text{act}^\Gamma/l_\text{p}^2\approx-0.6$ for both models, in approximate agreement with the theoretical value $-1/2$ from Eq.~\eqref{eq_cavobsGAMMA}. 
On the other hand, the offset, i.e., the adsorption $\Gamma(0)$ at a planar wall again differs between all approaches.

Regarding the different (model-dependent) values of the adsorption at zero curvature, we make the intriguing observation 
that they always equal the initial slope of the pressure in both theory and active simulations up to the factor $\Da$. 
Moreover, the factor $l_\text{p}\sqrt{\pi}/2$ occurs in all theoretical formulas for the active pressure $\p{W}$, Eq.~\eqref{eq_cavobs},
the adsorption $\Gamma$ (or effective surface tension), Eq.~\eqref{eq_cavobsGAMMA}, and the active surface tension $\sigma$, Eq.~\eqref{eq_cavobsSIGMA}, i.e., we have
\begin{equation}
m^p/l_\text{p} = \beta \sigma(0)/\Da \rho_0 l_\text{p} = -\Gamma(0) / \rho_0 l_\text{p}\,.
\label{eq_relationpg}
\end{equation}
Even more explicitly, the EPA results $\sigma(R^{-1})$ and $\p{W-}(R^{-1})$ for an active system confined to a cavity
are related by the same sum rule, Eq.~\eqref{eq_itsumrule}, as found for these quantities in equilibrium.
 In addition, an effective Gibbs adsorption theorem holds between the planar adsorption $\Gamma(0)$ and surface tension $\sigma(0)$, which is defined by replacing the thermal energy scale $\beta$ in Eq.~\eqref{eq_Gibbsadsorption} with the active one $\beta_\text{eff}$.
Assuming that such a relation exists in general, we also insert the
 (negative and rescaled) adsorption $-\Da\Gamma(R^{-1})$ into Eq.~\eqref{eq_itsumrule} and find again that this relation is fulfilled, irrespective of the difference at linear order in curvature from $\sigma(R^{-1})$.
The higher-order contributions for $\p{W+}(R^{-1})$, are, however, not recovered from such a sum rule in either way,
reflecting the issue that the behavior of active particles at an obstacle is more nonequilibrium-like in nature.
To examine this behavior for the active system in more detail, 
we also calculate the active surface tension $\sigma_\text{act}(0)$ at a planar hard wall, according to Eq.~\eqref{eq_gamma0}. 
Also for this quantity, we find a nice agreement with the initial slope, suggesting that Eq.~\eqref{eq_itsumrule} generally holds for an active ideal gas at a hard wall up to linear order in $R^{-1}$. Note that higher-order terms in $R^{-1}$ are difficult to determine accurately in our active simulations, due to large amounts of statistical noise.
All calculated coefficients for the different models are summarized in Table~\ref{tableSUM}.
 Before closing this section let us make two further comments.

There is an intriguing analogy to a passive ideal gas at a soft harmonic wall ($n\!=\!2$, see appendix~\ref{app_pots} for the specification of the potential).
For both a cavity and an obstacle, the exact formulas, 
\begin{align}
\frac{p(R^{-1})}{\rho_0}&=1-\frac{\sqrt{\pi}}{2}\frac{d}{\sqrt{\lambda}\,R}\,,\label{eq_pPASS}\\
\frac{\Gamma(R^{-1})}{\rho_0}&=\frac{\sqrt{\pi}}{2}\frac{d}{\sqrt{\lambda}}-\frac{1}{2}\frac{d^2}{\lambda R}=-\frac{\beta\sigma(R^{-1})}{\rho_0}\,,\label{eq_GammaPASS} %
\end{align}
 equal the results for an active ideal gas in a cavity upon identifying the two length scales $d/\sqrt{\lambda}$ and $l_\text{p}$.
Apparently, this analogy does not extend to the active (total~\cite{clarifyST}) surface tension which lacks the curvature term appearing in Eq.~\eqref{eq_GammaPASS}.
This leaves the impression that the adsorption in an active system exhibits more similarities to equilibrium than the surface tension, which is also suggested by the results in three dimensions, discussed in appendix~\ref{app_circular}.
Returning to the active simulation results in a cavity, which obviously display 
higher-order terms in $l_\text{p}/R$,
we realize that Eq.~\eqref{eq_itsumrule} no longer provides an accurate relation between the active pressure and adsorption beyond the initial slope.
Our observations at linear order thus identify an effective equilibrium regime \cite{fodor2016} for both AOUPs and ABPs.

As  anticipated from the significant deviations between the different approaches resulting from the behavior in the bulk,
 we find that all observations change dramatically 
when normalizing by the total particle number (in the bulk and adsorbed at the walls), which we elaborate in Appendix~\ref{app_norm}.
Most notably, the behavior of the two active models becomes much more consistent and agrees well with the EPA result.
In all cases, the pressure and adsorption at an obstacle are simply zero. 
For a very small cavity or highly persistent particles, such that all particles can be found at the wall, all results scale only with the local wall curvature $R^{-1}$.
However, this 
alternative normalization, comes at the cost of impairing the possibility to observe any relation reminiscent of an equilibrium sum rule, 
and the transition from a cavity to an obstacle is no longer smooth.
In a related study of a sinusoidal wall, where convex and concave regions naturally receive a unified normalization,
the difference between extremal pressures has been observed to be a linear function of the curvature~\cite{nikola2016}.
Since this obviously extends to walls with a noninfinitesimal curvature, 
the EPA can capture this observation only approximately, which we elaborate in Sec.~\ref{sec_sinus}.

\subsubsection{Structured walls \label{sec_sinus}}

Having demonstrated in Sec.~\ref{sec_circular} that the most sensible definition, $\p{W}$ in Eqs.~\eqref{eq_pW} and~\eqref{eq_pWofR}, of a wall pressure by the force exerted on the actual wall
is also the most appropriate one,
we now calculate the local pressure $\tilde{p}^\text{(W)}(y)$ on a structured wall with a modulation in $y$-direction,
which will also tell us more about the role of curvature in the EPA.
In principle, the formula from Eq.~\eqref{eq_pW} can be applied with the according potential $\Vext(x,y)$ to determine an expression for $\tilde{p}^\text{(W)}(y)$. 
However, this procedure becomes inconvenient for regions of positive geometrical curvature, as shown in appendix~\ref{app_curv}.

To efficiently study the general case of a structured (hard) wall with an arbitrary change in curvature, 
let us first note that, in the hard-wall limit, our result from Eq.~\eqref{eq_cavobsGAMMA} for the adsorption $\Gamma(R^{-1})$ is consistent with a more general expression
found in Ref.~\onlinecite{fily2017} as a function of the local curvature $\kappa$ of a nonuniform wall.
To see this, we simply have to identify
  $\kappa$ with the inverse radius $R^{-1}$ for a circular wall (notice the different convention for the sign of the curvature radius used here).
The derivation in Ref.~\onlinecite{fily2017} is based on the assumption that the confining potential is always normal to the wall structure and has a positive slope.
This is, strictly speaking, only justified in the hard-wall limit, which we discuss in appendix~\ref{app_curv}.
Adopting this strategy for the active pressure and generalizing it to positive values of the geometrical curvature, 
we find $p^\text{(W)}(\kappa)$ as a function of the (signed) local curvature 
in the form of Eq.~\eqref{eq_cavobs} with $R^{-1}\rightarrow\kappa$.
In other words, the pressure (and the adsorption) depend only locally on the curvature of the hard wall.

\begin{figure} [t] \centering
\includegraphics[width=0.425\textwidth] {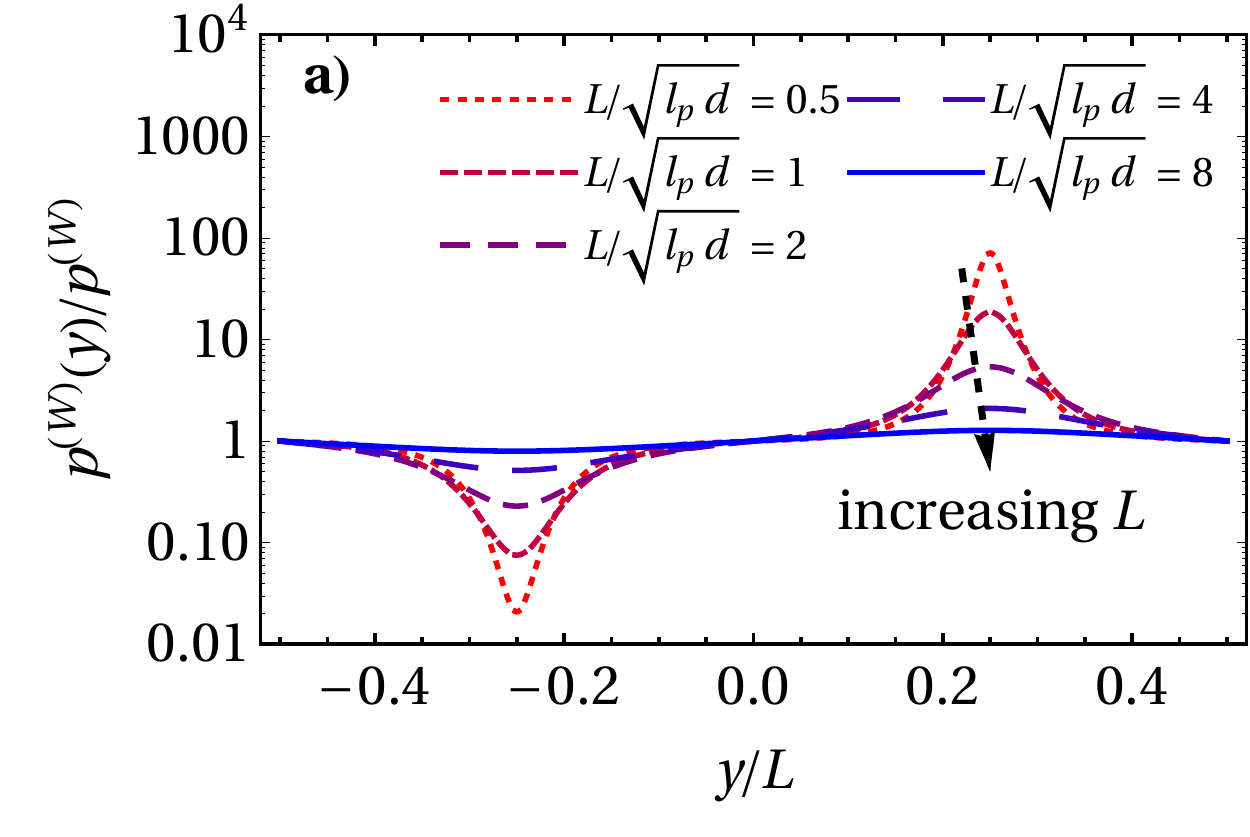}\\\vspace*{-0.2cm}
\includegraphics[width=0.425\textwidth] {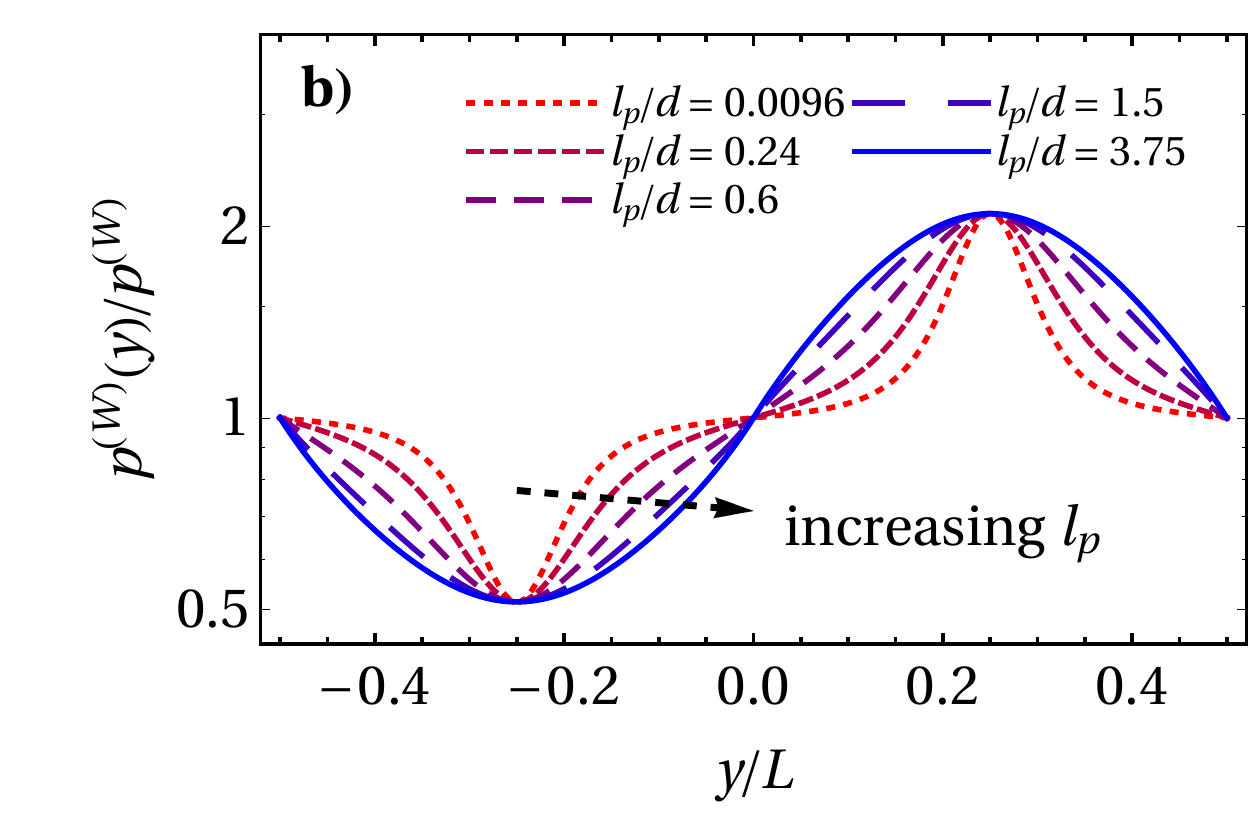}
\caption{Pressure $\tilde{p}^\text{(W)}(y)$ of an active ideal gas on a nearly hard sinusoidal wall of period $L$ parametrized by the coordinate $y$ (see text).
The pressure normalized by its bulk value depends explicitly on both $L$ and the persistence length $l_\text{p}$.
We compare the data for
\textbf{(a)} changing $L$ at constant persistence length $l_\text{p}=0.6d$  and \textbf{(b)} changing $l_\text{p}/d$ and adapting $L$ according to a fixed ratio $L/\sqrt{l_\text{p}d}=4$, so that the extreme values of the curvature remain constant. 
}
\label{fig_idXY}
\end{figure}

To demonstrate the implications of a pressure which only depends on the local curvature,
 we establish a connection to the simulations performed in Ref.~\onlinecite{nikola2016,sandford2018}
and consider the active pressure $\tilde{p}^\text{(W)}(y)$ on a sinusoidal wall of periodicity $L$
specified by the modulation function $M(y)=\sin(2\pi y/L)/2$ (the bulk can be found at $x<M(y)$). 
Employing the strategy described above, we define $\tilde{p}^\text{(W)}(y)=p^\text{(W)}(\kappa(y))$,
substituting the local curvature
\begin{align}\!\!\!
\kappa(y)= \frac{{d}\,M''(y)}{(1+{d^2}M'(y)^2)^\frac{3}{2}}=\frac{-\frac{{d}\,2\pi^2}{L^2}\sin(2\pi y/L)}{\left(1+\frac{{d^2}\pi^2\cos^2(2\pi y/L)}{L^2}\right)^\frac{3}{2}}
\end{align}
into Eq.~\eqref{eq_cavobs}.
In Fig.~\ref{fig_idXY} we illustrate the behavior of $\tilde{p}^\text{(W)}(y)/ \p{W}$ for different parameters. 
The overall qualitative picture is in nice agreement with the numerical expectation~\cite{nikola2016,sandford2018}
that the pressure becomes extremal at the apices of the modulation function,
with its maximum in the negatively curved region. 
Increasing $L$ at constant persistence length $l_\text{p}$ (Fig.~\ref{fig_idXY}a) or, vice versa,
decreasing $l_\text{p}$ at constant $L$ (not shown), the amplitude decreases.
Choosing $L/\sqrt{l_\text{p}}=\text{const}$, we can ensure that the maximal and minimal pressure remain independent of the activity (Fig.~\ref{fig_idXY}b).

However, there are two observations which highlight some underlying quantitative flaws of the theory.
First, the average pressure $\langle \p{W}\rangle\!=\!\int_0^L\upd y\,\tilde{p}^\text{(W)}(y)/L$ is always larger than the bulk value $\p{W}$ for the chosen periodic modulation,
since it becomes obvious from Fig.~\ref{fig_idC} and also Eq.~\eqref{eq_cavobs} that $\p{W}(-\kappa)\!-\!\p{W}\!\geq\!\p{W}\!-\!\p{W}(\kappa)$ for all $\kappa\!>\!0$ with an equality only in the planar limit $\kappa\rightarrow0$. 
Hence, only for sufficiently large $L$ or small $l_\text{p}$, Fig.~\ref{fig_idXY} illustrates that the ratio $\langle \p{W}\rangle/\p{W}$ consistently approaches 1,
which is an obvious result in both limits $L\rightarrow\infty$ of a flat wall and $l_\text{p}\rightarrow0$ of a passive system. 
Second, the theoretical pressure $\tilde{p}^\text{(W)}(nL/2)$ at the points of zero curvature is always equal to the bulk value $\p{W}$,
whereas simulations predict a smaller local pressure \cite{nikola2016,sandford2018}.
This also appears to be the reason for the first inconsistency.
The expected nonlocal dependence on the wall curvature makes sense when considering the active nature of the particle:
it will slide along the boundary until it detaches 
- either due to reorientation or due to a change in the structure of the wall.
For the given wall modulation, the change in curvature provides a way for particles to escape the flat part of the wall, 
lowering the density of particles at the wall (and hence the pressure) in the vicinity.
 Only in the limit of an infinite persistence length a local dependence on wall curvature can be expected \cite{fily2017}.

 \subsection{Interactions and wall softness}

\begin{figure} [t] \centering
\includegraphics[width=0.45\textwidth] {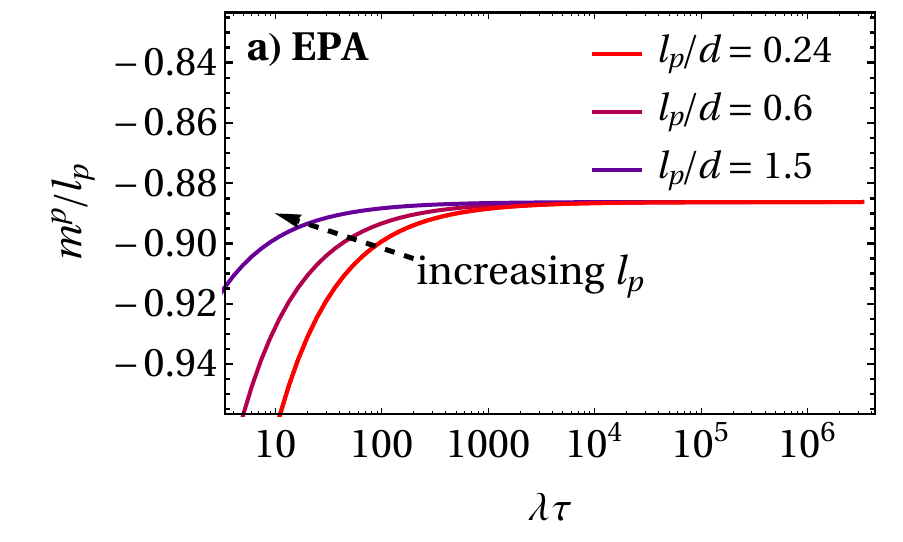} 
\includegraphics[width=0.45\textwidth] {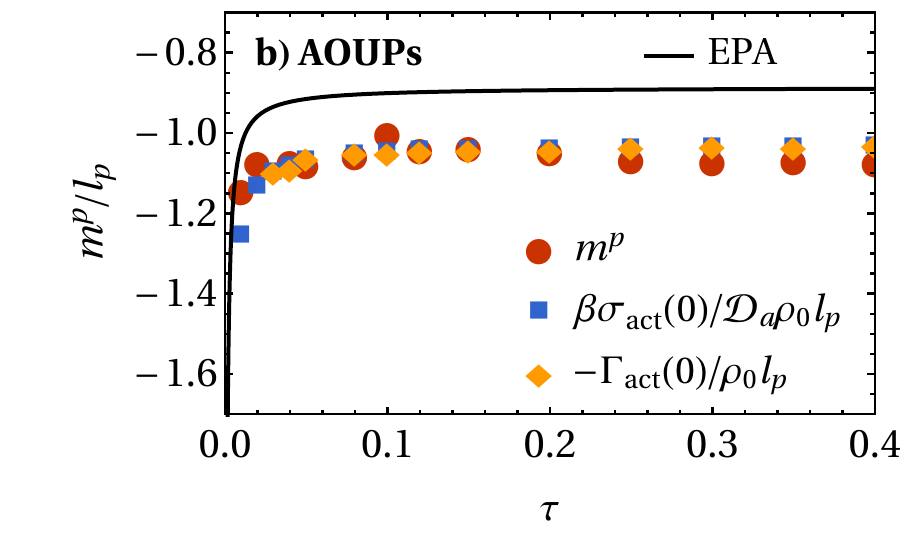} 
\includegraphics[width=0.45\textwidth] {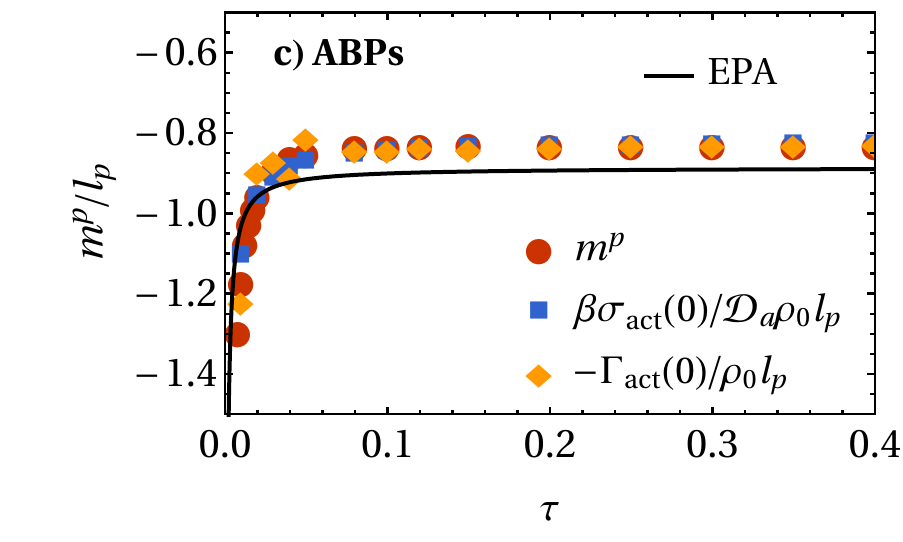} 
\caption{
Initial slope $m^p$ of the active pressure on a circular wall, as well as normalized adsorption $\Gamma(0)$ and active surface tension $\sigma(0)$ on a planar wall. 
We consider an ideal gas at a soft wall, with the potential specified in appendix~\ref{app_pots}.
Shown are \textbf{(a)} the identical theoretical results (EPA) for different persistence lengths $l_\text{p}$ as a function
 of the product $\lambda\tau$, where $\lambda$ is the softness parameter and $\tau$ the persistence time.
 Note that in all cases, Eq.~\eqref{eq_relationpg} is fulfilled.
In the hard wall limit, $\lambda\rightarrow\infty$, all curves approach $-\sqrt{\pi}/2$, compare, e.g., Eq.~\eqref{eq_cavobs}.
 We also compare, as a function of $\tau$, simulation results of all three quantities 
for \textbf{(b)} AOUPs and \textbf{(c)} ABPs to the EPA. 
Here we use the softness parameter $\lambda=3000$ and self-propulsion velocity $v_0 = 24  d / \tau_0$.
}
\label{fig_idCis}
\end{figure}

 Our final goal is now to continue the numerical study of interacting active and effective systems from Sec.~\ref{sec_planar}
 with a focus on curvature dependence. 
 We know from Sec.~\ref{sec_circular} that a proper comparison of different models is problematic for highly curved boundaries
 and from Sec.~\ref{sec_sinus} that the EPA does not capture a nonlocal dependence on the curvature. 
Thus, to only judge the quality of the effective pair interaction potential in a curved system,
we return to circular walls and restrict ourselves to the initial slope $m^p$, given by Eq.~\eqref{eq_IS}. 
Note that, while we focus in our simulations on the case of a circular cavity of (very large) radius $R$, 
the initial slope is expected to be the same for a circular cavity and a circular obstacle, and numerical tests for selected points confirm this. 
For systems with sufficiently low activity ($l_\text{p} \ll R$), the first-order result $m^p$ should still provide a good estimate for the curvature dependence of the wall pressure. 
In further contrast to the study in Sec.~\ref{sec_circular},
 it is necessary for the computer simulations of an interacting system to consider a slightly soft wall.
In general, this will be taken into account by choosing the finite value $\lambda=3000$ of the softness parameters in the interaction potentials, compare appendix~\ref{app_pots}.
 Also recall that the bulk formula $\pact{B}$ cannot be used to study the dependence on the wall curvature.

  \subsubsection{Active ideal gas at a soft circular wall \label{sec_softit}}

 As a first step we need to understand the role of the softness of the wall for an active ideal gas,
 which is recovered as the low-density limit of an interacting system.
  The softness parameter $\lambda$ of the wall potential (compare appendix~\ref{app_pots}) now provides an additional length scale and the results do not any more depend only on one universal argument.
  The theory suggests that the initial slope $m^p$ 
  depends explicitly on both the product of $\lambda$ with the persistence time $\tau$, as well as, the persistence length $l_\text{p}$,
even if we divide by $l_\text{p}$.
Only for very large values of $\lambda\tau$ all theoretical curves in Fig.~\ref{fig_idCis}a for different persistence lengths $l_\text{p}$ 
collapse on the same line, 
 i.e., the persistent limit is formally equivalent to the hard-wall limit (infinite $\lambda$).
 Note that the limit of $\tau\rightarrow0$ at fixed self-propulsion velocity (as shown in Fig. \ref{fig_idCis}) does not correspond to the limit of a passive system (where instead $\Da=1$ should be kept fixed).

The main point we wish to make here concerns the relation between pressure and surface tension (or adsorption). Explicitly, 
in generalization of the study from Sec.~\ref{sec_circular}, 
 we find for a soft cavity
\begin{align}
\frac{\p{W-}(R^{-1})}{\Da\rho_0}&=1-\frac{\sqrt{\pi}}{2}\,\frac{\sqrt{\Da(1+2\lambda\tau)}}{\sqrt{\lambda}}\,\frac{d}{R}\,,\label{eq_pSOFT}\\
\frac{\Gamma^{(-)}(R^{-1})}{\rho_0}&=\frac{\sqrt{\pi}}{2}\,\frac{\sqrt{\Da(1+2\lambda\tau)}}{\sqrt{\lambda}}\,d-\frac{1}{2}\frac{\Da(1+2\lambda\tau)}{\lambda}\frac{d^2}{R}\,,\label{eq_GammaSOFT}\\
\frac{\sigma^{(-)}(R^{-1})}{\Da\rho_0}&=\frac{\sqrt{\pi}}{2}\,\frac{\sqrt{\Da(1+2\lambda\tau)}}{\sqrt{\lambda}}\,d-\frac{1}{2}\frac{\Da}{\lambda}\frac{d^2}{R}\,.\label{eq_SigmaSOFT}
\end{align}
All expressions are still linear in $R^{-1}$ and reduce to Eqs.~\eqref{eq_cavobs},~\eqref{eq_cavobsGAMMA} and~\eqref{eq_cavobsSIGMA} in the hard-wall limit, $\lambda\rightarrow\infty$,
as well as to Eq.~\eqref{eq_pPASS} and~\eqref{eq_GammaPASS} in the passive limit, $\Da=1$ and $\tau=0$.
Apparently, the theory provides the same analytic coefficients for the initial slope $m^p$ of the pressure, the planar surface tension $\sigma(0)$ and the adsorption $\Gamma(0)$ at a planar wall.
 Comparing Eq.~\eqref{eq_GammaSOFT} with Eq.~\eqref{eq_SigmaSOFT}, we notice that the deviation from an the effective Gibbs adsorption theorem, compare Eq.~\eqref{eq_Gibbsadsorption}, at linear order in the curvature is by a term which is independent of the wall softness and solely due to activity, i.e., it (necessarily) vanishes in the passive limit.
These theoretical results are again nicely confirmed by active simulations, which we compare in Figs.~\ref{fig_idCis}b and c for a fixed wall potential.
Within numerical accuracy Eq.~\eqref{eq_relationpg} holds for any persistence time
 in all models considered.
These observations suggest that the sum rules discussed in Sec.~\ref{sec_circular} are still at work up to linear order when we allow for a finite wall softness.
Also note that $m^p$ is again independent of the route to calculate the pressure ($\p{W}$ or $\p{T}$).

In the regime where we study the interacting system ($\lambda=3000$ and $\tau<0.05$), the results of all models deviate noticeably from the hard-wall limit.  
For both ABPs and AOUPs, there seems to be a (weak) effect of the strength of the wall potential $\lambda$ (without the factor $\tau$). 
This is potentially related to the interplay between the effective interaction range of the wall and the persistence length. 
Up to the offset between the different models already observed for a hard wall, all curves in Fig.~\ref{fig_idCis}b and c are qualitatively similar to the theoretical result.
However, the slightly different slope of $m^p$ in the different models gives rise to a spurious point where the theory and simulations are in perfect agreement.
The corresponding parameter $\tau=0.025$ for ABPs appears to be a convenient choice to study the influence of interactions,
 although the agreement is rather coincidental.
Note, however, that the differences observed for such small $\tau$ are insignificant, 
since we normalize here by the persistence length which becomes equally small in this region.

  \subsubsection{Interacting particles at a curved surface}

\begin{figure} [t] \centering
\includegraphics[width=0.475\textwidth] {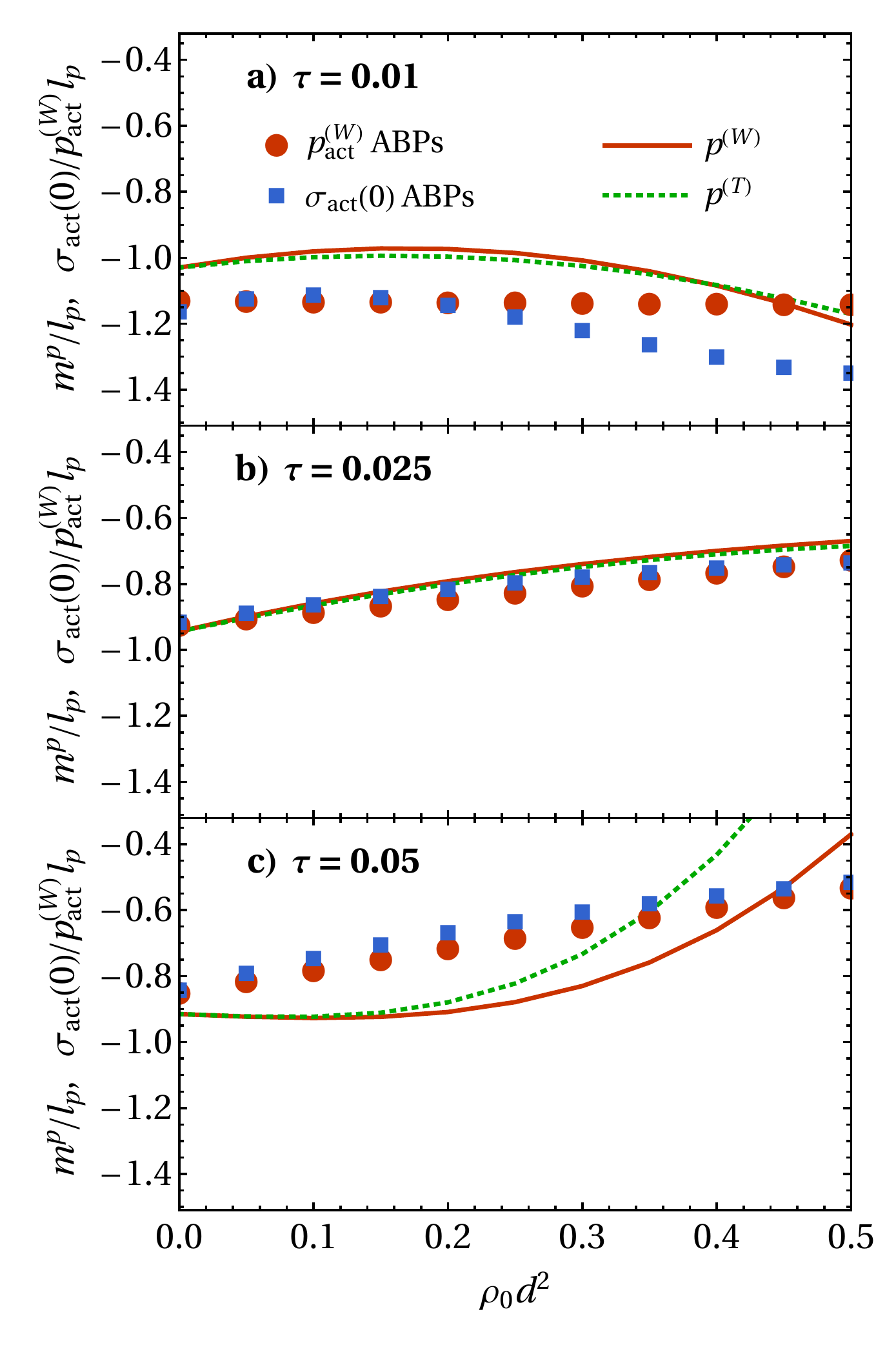}
\caption{
Initial slope $m^p$, given by Eq.~\eqref{eq_IS}, of the dependence of the active pressure $p$ on the wall curvature $R^{-1}$ in an interacting system of ABPs, as a function of density $\rho_0$.
We compare the wall pressure $\pact{W}$, measured directly in an active system, 
and the bare and effective wall pressure $\p{W}$ and $\p{T}$, measured in the corresponding passive system according to Eq.~\eqref{eq_pWofR} and Eq.~\eqref{eq_pTofR}, respectively.
Moreover, we show the normalized active surface tension $\sigma_\mathrm{act}(0)$ measured for ABPs at a planar wall.
We consider three different persistence times
{\bf (a)}~$\tau \!=\! 0.01$, {\bf (b)}~$\tau \!=\! 0.025$ and {\bf (c)}~$\tau \!=\! 0.05$
at fixed self-propulsion speed $v_0 \!=\! 24 d/\tau_0$ and use the same scale on all axes for a better comparison of the influence of activity on the density dependence. 
Note that the points corresponding to ABPs are based on fits to the simulation data.
The $\tau$-dependent offset at $\rho_0\!=\!0$, described in Sec.~\ref{sec_softit}, should not be mistaken for an optimal agreement at intermediate activity.
}
\label{fig_p1}
\end{figure}

We now compare the curvature dependence in interacting active and effective systems, where we focus our attention on ABPs only.
In Fig.~\ref{fig_p1} we show the initial slope $m^p$ as a function of the density for different activity parameters, as well as the active surface tension at a flat wall $\sigma_\mathrm{act}(0)$.
For $\rho_0=0$, the result is equal to that of an ideal-gas for the corresponding parameters, which explains the offset between the curves for ABPs and the EPA. 
 Despite this systematic deviation, the density dependence of the EPA result $p^\text{(W)}(R^{-1})$ shows adequate agreement with the wall pressure of the active system, demonstrating that the EPA correctly captures the curvature dependence of the wall pressure, as long as the curvature is not too high.
 In particular, the interplay between interactions, activity and curvature, which we discuss at the end of this section, can be qualitatively reproduced.

At finite densities, the normalized initial slope $m/l_\text{p}$ explicitly depends on both the chosen persistence time and persistence length, since there is an additional length scale given by the particle size.
 Interestingly, the initial slope $m^p$ of the curvature dependence for both theoretical pressures $p^\text{(W)}$ and $\p{T}$ is identical within our error bars at low activity, even though the predicted pressures are not the same, cf., Fig.~\ref{fig_p0}.
This can be understood from the fact that the curvature-dependence of the wall pressure is mainly caused by the variation in particle density at the wall, i.e., the planar adsorption, and both pressures are based on the same density profile. 
  However, at higher activity ($\tau = 0.05$), we begin to observe deviations between the two.
 
  For the investigated activities, the active surface tension $\sigma_\mathrm{act}(0)$ again matches the linear effect of curvature on the pressure $m^p$. The largest deviation is seen at weak activity, where accurate determination of both the surface tension and the slope of the curvature-dependence of the pressure are hard to resolve accurately. Our estimation of  the statistical errors inherent in our numerical data  suggests that within our accuracy, the relation between surface tension and pressure on a curved wall is maintained even for interacting active systems. Note, however, that due to the necessity to fit our data and extrapolate to the limit of large cavities, our statistical errors in the surface tension are rather large, especially for small $\tau$, where the effects we measure are weak. We estimate that error bars in $\sigma_\mathrm{act}$ are on the order of 25\%, 15\% and 10\% for persistence times $\tau = 0.01, 0.025$ and $0.05$, respectively.

From our simulations we observe that for high activity
the effect of curvature on the pressure decreases significantly in a dense system,
which becomes apparent from the decreasing absolute value of the initial slope in Figs.~\ref{fig_p1}b and c for increasing density. 
A possible explanation for this stems from the escape mechanism of a trapped particle. 
In order for a particle to move away from any wall, it has to rotate its swimming direction away from its normal vector. 
A negative curvature (cavity) hinders this, as during this process the particle will slide along the wall towards the point where its swimming direction points towards the wall again. 
However, if the particle encounters another particle during this process, this sliding is inhibited, facilitating wall escape. 
Hence, if the density near the wall is high enough that the particles are likely to collide before they reorient, the effect of curvature is diminished. 
This is indeed expected to occur at high densities and strong activity. 
For lower activity, this trend competes with the (passive) surface tension, which tends to increase with number density,
so that in the case shown Fig.~\ref{fig_p1}a the initial slope remains nearly constant. The theory approximately reflects this behavior.

\section{Conclusions \label{sec_conclusions}}

In conclusion, we have shown that a mapping to equilibrium, which allows for the definition of effective interaction potentials, i.e., the EPA,
is a helpful tool to understand the mechanical pressure in an active system, measured through the mechanical force on a wall, i.e., by calculating the one-body density profile.
This is still true if the wall is moderately curved. 
For an interacting system we demonstrated that also the behavior of the density-dependent corrections to the pressure at linear order in the curvature can be nicely captured by our theory.
In general, our results help to better asses the range of validity of equilibrium mappings in higher dimensions.

The observations in a system with planar symmetry allow for a conclusion which has an intriguing analogy to a fundamental problem in equilibrium liquid-state theory \cite{archerevans2017}:
the artifacts of an approximate theory are much less severe on the one-body (density profile) than on the two-body (pair correlations) level.
Explicitly, the example considered in Ref.~\onlinecite{archerevans2017} is the mean-field theory,
within which the radial distribution determined from the density around a test particle outperforms 
the version obtained by taking two functional derivatives of their approximate free energy.
In our case, the approximation to be judged is the EPA with its equilibrium mapping and subsequent definition of pairwise effective interactions.
With this analogy in mind, one can expect that the prediction of phase transitions \cite{faragebrader2015} or wetting profiles \cite{wittmannbrader2016} should be more robust
than one might expect from the partly strong deviations of the radial distribution from the true ABPs result \cite{SpeckCRIT}.
The calculation of the pressure performed in the present work is consistent with this argument,
since we have established that the wall pressure, Eq.~\eqref{eq_pW}, obtained from the EPA, is much more accurate than the bulk formula, Eq.~\eqref{eq_pV}, which contains implicit approximations on the level of pair correlations, even when evaluated with the exact reference input from active simulations.
The active simulation results for the pressure measured in bulk and at a flat wall are, of course, equivalent.

To judge the accuracy of the multidimensional theory, 
it is not sufficient to argue about the persistence time alone. 
We have seen that pressure, adsorption and surface tension of an active ideal gas
rather depend universally on the product of persistence length and local wall curvature,
which is also found for active simulations at a hard wall with constant curvature.
Up to a model-dependent offset, we find in this special case an excellent agreement at leading order.
In practice, this means that one may equally consider a system with infinitesimal curvature and finite activity or one with finite curvature and infinitesimal activity.
In this respect, our simulation results
 both quantify the differences between the distribution functions of ABPs and AOUPs \cite{szamel2014,Das2018}
and demonstrate a certain universality of both models in the near-white-noise regime.
As we will further elaborate in the following two paragraphs, we conclude for the EPA that one should ideally consider a system with both low activity and small absolute values of the geometrical (and/or potential~\cite{sharma2017}) curvature to make reliable predictions.

While earlier studies were limited to a negative geometrical curvature, where the effective diffusion is positive,
we employed a modification of the EPA, which, despite its empirical nature,
allows us to reasonably extend the calculation to regions with positive geometrical curvature.
The insights we obtained from ideal active particles at obstacles with a nonzero curvature
are also important for bulk systems of interacting particles,
which becomes obvious from drawing the analogy to the interaction with another particle.
In this case the wall curvature is represented by the particle radius.
This means that the behavior of the effective pair potentials
overestimating the attraction between two active particles \cite{activePAIR,faragebrader2015}
has the same origin as the deviation between the ideal pressure and the simulation results for the different models at a given curvature of the obstacle.
Since the size of the particle provides a fix length scale,
it becomes apparent that we can only obtain accurate results in an interacting system in the small-activity regime.

For a cavity, we found that it is not as easy as it appears from some statements in the literature to argue about the persistent limit in the present application of the equilibrium mappings.
The reason is a model-dependent exponent entering the chosen normalization factors through the depleted bulk density in a cavity.
When normalizing the total number of particles, as in appendix~\ref{app_norm}, we recover
a universal persistent limit of the adsorption in the theory and in simulations of ABPs and OUPs \cite{fily2017}.
The theoretical description for the pressure exactly matches the AOUPs pressure,
where we recall that the theory is designed to mimic the behavior of this model and not that of ABPs. 
In this sense, the statement that the theory becomes exact for infinite persistence time is justified,
albeit we should add the restriction that this is only the case if we neglect the subtle influence of an ill-defined bulk in the active systems.
 This makes sense if we remind ourselves that the (vanishing) external field in the bulk region is not strictly convex.
Obviously, the passive limit with vanishing persistence length is always exact, as well as, the (ideal-gas) pressure measured in the flat-wall limit of zero curvature.

Our work will guide the way to test alternative theories \cite{fily2018,caprini2018}
or to improve upon the multidimensional Fox approach and the Unified Colored Noise Approximation.
 Regarding these current equilibrium mappings, one has to examine in more detail the central building block, namely the effective diffusion tensor defined in Eq.~\eqref{eq_Deff}.
In particular, the relatively simple functional dependence on the interaction force does not admit higher-order terms in the curvature expansion for a cavity, away from the persistent limit.
Moreover, the inability to reproduce the behavior of the pressure at a structured wall with a modulating curvature
indicates that also higher-order derivatives of the interaction force should be accounted for.
In other words, in the present theory
a particle only ``sees'' the curvature rather than the change of curvature.
This point will be addressed in more detail in a future publication.

Despite the discussed limitations, our results show that equilibrium simulations using effective potentials can 
provide strikingly good predictions for the behavior of the pressure in both flat-wall and curved-wall geometries, as long as the activity is sufficiently small. 
Note that the numerical and analytical tools used here can be easily employed in three dimensions as well, 
where the wall pressure shows a similar curvature dependence as in two-dimensional systems. 
The possibility of studying a passive equivalent of an active system vastly simplifies the study of active systems in simulations. 
In particular, the simulation of equilibrium systems allows for the application of computational techniques that are not valid in nonequilibrium systems, including Monte Carlo simulation,
biased sampling schemes, free-energy calculations, and simulations in different ensembles.
Therefore, the EPA is potentially a strong tool in improving our understanding of active systems.

Most notably, as motivated by the analytic results of the theory,
 we identified and confirmed by active simulation two nonequilibrium relations reminiscent of equilibrium sum rules for an ideal active gas.
First, the excess adsorption at a planar wall can be determined, using the Gibbs adsorption theorem, from the integrated surface excess pressure (total surface tension) up to an activity-dependent factor, which is given by the well-known effective temperature.
Introducing a finite curvature, the theory predicts a deviation from this behavior, thereby revealing an additional direct contribution to the adsorption due to activity.
The planar coefficients of both quantities further match the slope of the curvature-dependent pressure at zero curvature,
indicating that another sum rule holds between pressure and surface tension 
or (negative and rescaled) adsorption if the wall is infinitesimally curved.
 Up to our numerical accuracy, our simulations confirm the first relation even for interacting active systems, which is not {\it a priori} obvious due to the presence of nonequilibrium correlations between particles.

 Our findings suggests that, in the small-curvature limit (and similarly in the low-activity limit), 
there exists a quantity with the same properties (at an effective temperature) as the chemical potential in equilibrium. 
Beyond this equilibrium-like regime the considered sum rules might provide a new way to introduce an active chemical potential or to cross-check other attempts to define such a quantity \cite{paliwal2018chemical,meer2016chempot}.
Hence, understanding the exact relation between pressure, adsorption and surface tension in an active system---if only for an ideal gas---provides an important further step in the direction of understanding the thermodynamics of active matter.
 This task could be achieved by generalizing the analytic solutions for AOUPs in a harmonic trap \cite{Das2018}
to the shifted potentials of finite radius considered here, or employing refined approximations \cite{fily2018,caprini2018}.
 Further theoretical efforts should clarify in how far the one-body swim contribution \cite{takatori2014} 
to the total surface tension can be expressed terms of the adsorption, 
analog to the ideal contribution (in the mechanical definition) for an equilibrium fluid \cite{kirkwoodbuff}.
A more detailed numerical study including higher-order terms in the curvature expansions of 
 the mechanical surface quantities would offer deeper insights in the limitations to the applicability of equilibrium sum rules and aid future theoretical investigations.

\section*{Acknowledgements}

The authors would like to thank Hartmut L\"owen for initiating this collaboration and a critical reading of the manuscript.
 Bob Evans helped us a lot by clarifying what exactly should be called a surface tension at an external wall.
Helpful discussions with Umberto Marini Bettolo Marconi and Abhinav Sharma are gratefully acknowledged.
R.\ Wittmann and J.\ M.\ Brader acknowledge funding provided by the Swiss National Science Foundation under Project No.\ 169073.

\appendix

\section{Simulation details  \label{app_sim}}

The active Brownian systems (ABPs) were simulated using overdamped dynamics simulations in two spatial dimensions, following the equations of motion
\begin{eqnarray}
\dot{\mathbf{r}}_i &=& 
\gamma^{-1}\bvec{F}_i + \gamma^{-1} f_0 \, \mathbf{n}_i  \nonumber\\
\dot{\phi}_i       &=& \sqrt{2 D_\text{r}} \,\eta_i(t). \label{eq:eom}
\end{eqnarray}
for the positions $\mathbf{r}_i$ and orientation angles $\phi_i$.
Here, $\bvec{F}_i$ are the conservative forces on particle $i$ resulting from the bare pairwise interactions and wall interactions, defined in appendix~\ref{app_pots}.
Additionally, $\gamma$ is the friction coefficient of a single swimmer,
$\mathbf{n}_i = \{\cos \phi_i, \sin \phi_i\}$ are unit vectors representing the particle orientations,
$D_\text{r}$ is the rotational Brownian diffusion coefficient
and $f_0$ is the constant absolute force which describes the self-propelled movement.
Finally, $\eta_i(t)$ represents a delta-correlated stochastic noise term with zero mean and unit standard deviation.
Note that the particles do not undergo translational (Brownian) diffusion.
The parameters that enter the effective steady-state condition, Eq.~\eqref{eq_FP} of the main text, via the effective diffusion tensor 
are the self-propulsion velocity $v_0=\gamma^{-1}f_0$ and the reorientation time $\taua=D_\text{r}^{-1}$.

For the simulations of the active Ornstein-Uhlenbeck model (AOUPs), we follow the same strategy, but use the following equation of motion:
\begin{equation}
\dot{\mathbf{r}}_i = \gamma^{-1}\bvec{F}_i + \boldsymbol{\chi}_i\,,
\label{eq_OUPs}
\end{equation}
where the stationary AOUPs $\boldsymbol{\chi}_i(t)$ evolve in time according to
\begin{equation}
\dot{\boldsymbol{\chi}}_i(t)=-\frac{\boldsymbol{\chi}_i(t)}{\taua}+\frac{\boldsymbol{\xi}_i(t)}{\taua}\,.
\label{eq_OUPsDEF}
\end{equation}
We choose here the same timescale $\taua$ as for rotational Brownian motion
and consider a stochastic noise vector $\boldsymbol{\xi}_i$ with zero mean and $\langle\boldsymbol{\xi}_i(t)\boldsymbol{\xi}_j(t')\rangle\!=\!2D_\text{a}\boldsymbol{1}\delta_{ij}\delta(t-t')$, where $D_\text{a}=v_0^2\taua/2$.
With this choice, the AOUPs $\gamma\boldsymbol{\chi}_i$ and the self-propulsion force $f_0 \, \mathbf{n}_i $ of ABPs have the same standard deviation \cite{faragebrader2015}.

For simulations of single-particle systems with hard walls, we use the same event-driven approach as in Ref. \onlinecite{smallenburg2015}. In these simulations, we update the propulsion vector of the particle at fixed time intervals set by the integration time step, but within each time step resolve the trajectory of the particle exactly by taking into account collisions with the walls, again assuming overdamped dynamics.

Simulations were performed either in bulk (no walls), in the presence of two flat walls, in the presence of a circular obstacle with radius $R$, or inside a circular cavity with radius $R$. 
All simulations used a fixed time step of $\delta t = 10^{-4} \gamma^{-1}$.
Pressures and pressure tensors were determined by directly measuring the force per unit length exerted by the particles on the walls in the system,
or (for the bulk case) via the virial-based approach introduced by Winkler {\it et al.} \cite{winkler2015}.
Note that the bulk pressure and the flat-wall pressure coincide in sufficiently large systems. 
Moreover, we calculated the bulk pressure from the radial distribution function according to Eq.~\eqref{eq_pV} of the main text.
For the bulk, flat-wall, and curved-wall systems with positive geometrical curvature (circular obstacle), we used a fixed number of particles $N = 2000$, while in the systems with negatively curved walls (cavities),
the number of particles was adapted to the confining container. For both cavities and obstacles, we considered curvature radii between $R/d = 5$ and $R/d = 30$.

The passive systems, using the effective interaction potentials defined in appendix~\ref{app_pots}, 
were studied by means of standard Monte Carlo simulations in the canonical ensemble
for the same system geometries as for the active systems. 
The pressure was measured either via the radial distribution (for bulk simulations) or by directly measuring the force per unit length exerted by the particles on the wall,
see main text for the explicit definition of all expressions for the pressure considered.

\section{Interaction potentials \label{app_pots}}

In this appendix, we specify the interaction potentials employed in the two-dimensional calculations throughout the manuscript.
We model the bare particle-wall interactions by the power-law potentials~\cite{fily2017}  
\begin{align}
  \beta \Vext(x)&= \lambda\left(\frac{2 x}{d}\right)^n\Theta(-x)
\end{align}
for a planar wall
and  
\begin{align}
  \beta \Vext^-(r)&= \lambda\left(\frac{2(r-R)}{d}\right)^n \Theta(r-R)\,,\label{eq_v0m}\\
  \beta \Vext^+(r)&= \lambda\left(\frac{2(r-R)}{d}\right)^n\Theta(R-r)\label{eq_v0p}
\end{align}
for a circular walls of radius $R$, which corresponds to the negative curvature radius of a cavity $(-)$
and the positive curvature radius of an obstacle $(+)$, respectively.
Moreover, $\Theta(x)$ denotes the Heaviside step function, $\lambda$ is a parameter controlling the softness of the wall and $d$ provides the length scale of the particles.
For the pair interaction between the particles, we assume 
\begin{align}
   \beta u(r)&= \lambda\left(\frac{r-\tilde{R}}{d}\right)^n\Theta(\tilde{R}-r)\Theta(r). \label{eq_vbare}
\end{align}
with fixed particle radius $\tilde{R}$.

Taking the limit $\lambda\rightarrow\infty$ \textit{after} having calculated the property of interest (effective potential, pressure,...) 
we obtain the results corresponding to a hard wall. 
We made sure that the result in the hard-wall limit is independent of the exponent, so, for simplicity, we choose $n=2$ in the analytic study of the ideal gas at the curved surfaces.
In all our numerical studies, we choose the exponent $n=4$ and the parameters $\tilde{R}=2^{1/6}d$ and $\lambda=3000$ to obtain hard-core-like potentials.

 Solving Eq.~\eqref{eq_FP} in a radial geometry, we define the effective external fields
\begin{align}
 \!\!\!\!\!\!\!\!\!\!\!\!\!\!\!\! \beta \Vext^\text{eff-}(r)&=\beta\frac{ \Vext+\frac{{\ttau} (\Vext')^{2}}{2}}{\Da}-\ln\!\Big(\Big(1+{\ttau} \frac{\Vext'}{r}\Big)\big(1+{\ttau} 
  \Vext''\big)\Big), \label{eq_ve}\!\!\!\!\\
  \beta \Vext^\text{eff+}(r)&=\beta\frac{ \Vext+\frac{{\ttau} (\Vext')^{2}}{2}}{\Da}-\ln\!\Bigg(\frac{1+{\ttau}\Vext''}{1-{\ttau} \frac{\Vext'}{r}}\Bigg) \ \ \ \ \label{eq_vit}
\end{align}
for a cavity (positive slope of $\Vext(r)$) and an obstacle (negative slope of $\Vext(r)$), respectively.
To facilitate the notation, we omitted the arguments on the right-hand-side and have defined $\Vext'=\partial\Vext/\partial r$, $\Vext''=\partial^2\Vext/\partial r^2$ and $\ttau=\taua/\gamma=\beta\tau d^2$.
The formula for $\Vext^\text{eff+}(r)$ in Eq.~\eqref{eq_vit} for an obstacle has been modified compared to $\Vext^\text{eff-}(r)$ for a cavity
according to the inverse-$\tau$ approximation ensuring physically consistent results \cite{activePAIR}.
This empirical correction becomes necessary due to the negative slope (or positive geometrical curvature),
which we discuss in more detail in appendix~\ref{app_curv}.

Both expressions, Eq.~\eqref{eq_ve} and Eq.~\eqref{eq_vit}, for the effective external potential have the same linear-order term of an expansion in terms of $\tau$ and also $R^{-1}$.
Therefore, they become formally equivalent in the planar limit (infinite curvature radius),
yielding
\begin{align}
  \beta \Vext^\text{eff}(x)&=\beta\frac{ \Vext+\frac{{\ttau} (\Vext')^{2}}{2}}{\Da}-\ln\!\Big(1+{\ttau} \Vext''\Big),
  \label{eq_vpl}
\end{align}
with the Cartesian coordinate $x$. 
Finally, we define the effective pair potentials
\begin{align}
   \beta u^\text{eff}=\beta\frac{ u+{\ttau} (u')^{2}}{\Da}-\ln\!\Bigg(\frac{1+{2\ttau} u''}{1-{2\ttau} \frac{u'}{r}}\Bigg)
   \label{eq_veff}
\end{align}
in analogy to Eq.~\eqref{eq_vit}, since one particle formally acts as an obstacle to its neighbors.

\section{Geometrical curvature and slope of the potential \label{app_curv}}

The effective potentials specified in appendix~\ref{app_pots} can be defined in terms of the  
Eigenvalues of the effective diffusion tensor from Eq.~\eqref{eq_Deff} of the main text \cite{activePAIR}.
Restricting ourselves to a radial geometry for the moment,
we find the two Eigenvalues
\begin{align}
  \!\!\!  \! E_1(r)=\Da\left(1+{\ttau} \frac{\Vext'}{r}\right)^{-1},\  E_2(r)=\Da\left(1+{\ttau} \Vext''\right)^{-1}
\end{align}
compare, e.g., the expressions in the logarithm in Eq.~\eqref{eq_ve}.
The second Eigenvalue $E_2$ explicitly depends on the second derivative of $\Vext(r)$, i.e., the potential curvature.
It is always positive for the potentials considered in this work.
The first Eigenvalue $E_1$, however, explicitly depends on the first derivative (slope) of $\Vext(r)$,
which becomes negative at an obstacle.
To avoid the imminent unphysical divergence, we empirically set $E_1(r)=\Da(1-{\ttau}\Vext'/r)$
in this case \cite{activePAIR}.
Practically, this inverse-$\tau$ approximation restores the correct trend of the term linear in $\tau$ of $E_1(r)$ to increase monotonically
with increasingly negative slope of the potential.

The problem occurring in the theory for potentials with a negative slope is of the same nature as that for a positive geometrical curvature.
Indeed, these two properties are formally equivalent in the hard-wall limit, where the curvature radius radius is unambiguously defined as $\pm R$.
Since the potentials $\Vext^{\mp}(r)$ in Eqs.~\eqref{eq_v0m} and~\eqref{eq_v0p} only depend on the difference $r-R$, 
we first substitute $r\rightarrow \pm s+R$ to eliminate the dependence on $R$ and to unify the two different formulas for a cavity and an obstacle if the exponent $n$ is even.
The slope of 
\begin{align}
 \tilde{\Vext}(s)=\lambda\left(\frac{2s}{d}\right)^n\Theta(s)
\end{align}
is then always positive in the new coordinates,
which can be interpreted by choosing the $s$-axis always parallel to the surface normal.
  Then, we argue that, for a nearly hard wall, the value of $s$ is essentially zero and can be neglected when compared to the radius, i.e. $s\ll R$.
Finally, we identify the radius $R$ with the signed inverse curvature, i.e., $R\rightarrow \kappa^{-1}$ for an obstacle ($\kappa>0$)
and $R\rightarrow-\kappa^{-1}$ for a cavity ($\kappa<0$).
The result for the Eigenvalues is
\begin{align}
   \!\!\!  \! E_1(s)=\Da\left(1-{\ttau}\kappa\tilde{\Vext}'\right)^{-1},\  E_2(r)=\Da\left(1+{\ttau} \tilde{\Vext}''\right)^{-1},
   \label{eq_E12kappa}
\end{align}
where the sign of the curvature now takes the role of the sign of the slope.
Thus we have established that geometrical curvature and potential slope are equivalent (mind the sign convention)
and thereby reproduced the results of the entirely geometrical derivation in Ref.~\onlinecite{fily2017}.
The divergence of $E_1$ now can occur only for positive values of $\kappa$, i.e., only for an obstacle.
Employing the inverse-$\tau$ approximation amounts to set  $E_1(r)=\Da(1+{\ttau}\kappa\Vext')$ and has the same effect as in radial geometry.
In a planar geometry, the Eigenvalue $E_1$ is constant and does not contribute to Eq.~\eqref{eq_vpl},
which becomes directly apparent from the curvature representation in Eq.~\eqref{eq_E12kappa}, setting $\kappa=0$ for a flat wall.

Finally, we note that the interpretation of the coordinate $s$ to be always perpendicular to the wall
allows us to easily generalize the theory to the case of a hard modulating wall by substituting $\kappa\rightarrow\kappa(y)$ in the Eigenvalues from Eq.~\eqref{eq_E12kappa},
where $\kappa(y)$ is the local curvature at the space-fixed coordinate $y$ parallel to the wall.
The advantage of this approach is that there is no ambiguity in how to correct the effective potential in the regions with positive curvature,
in contrast to a potential $\Vext(x,y)$ with the soft wall potential increasing along the space-fixed $x$ coordinate,
where the Eigenvalues of $E_1(x,y)$ and $E_2(x,y)$ would depend on both coordinates simultaneously.
While being more difficult to be tackled in theory, such a setup is the physically more sensible one if the particle-wall interaction is soft,
which is necessarily the case in computer simulations~\cite{nikola2016,sandford2018}.

\section{
Ideal gas at a hard circular wall in three dimensions
\label{app_circular}}

Here, we briefly extend the theoretical investigation from Sec.~\ref{sec_circular} to three spatial dimensions,
i.e., we study an ideal active gas at a spherical hard wall.
All parameters, except for the active diffusivity $\Da=l_\text{p}v_0\tau_0/(3d^2)$ are defined in the same way as in two dimensions.
Focusing on the case of a cavity for simplicity, we find
in generalization of Eqs.~\eqref{eq_cavobs} and~\eqref{eq_cavobsGAMMA} the pressure
\begin{align}
\frac{p^\text{(W-)}(R^{-1})}{\Da\rho_0}&=1-\frac{\sqrt{6\pi}}{3}\frac{l_\text{p}}{R}+\frac{2}{3}\frac{l_\text{p}^2}{R^2}\,,
\label{eq_cavobs3d}
\end{align}
and the adsorption
\begin{align}
\frac{\Gamma^{(-)}(R^{-1})}{\rho_0}&=\frac{\sqrt{6\pi}}{6}l_\text{p}-\frac{2}{3}\frac{l_\text{p}^2}{R}+\frac{\sqrt{6\pi}}{18}\frac{l_\text{p}^3}{R^2}\,,
\label{eq_cavobsGAMMA3d}
\end{align}
respectively.
Compared to their two-dimensional analogs, both formulas contain an additional term quadratic in $R^{-1}$, representing the Gaussian curvature of the wall \cite{fily2017}.
The active surface tension
\begin{align}
 \frac{\beta\sigma}{\rho_0}=-\Da\frac{\sqrt{6\pi}}{6}l_\text{p}+\frac{1}{3}\frac{l_\text{p}^2}{R}
 \label{eq_cavobsSIGMA3d}
\end{align}
is now also curvature dependent, in contrast to Eq.~\eqref{eq_cavobsSIGMA}.
However, it does not depend on the full shape of the wall in general, i.e., only the mean curvature contributes. 

Finally, the three-dimensional equivalent of the equilibrium sum rule from Eq.~\eqref{eq_itsumrule} reads \cite{sumrules}
\begin{align}
p_\text{eq}(R^{-1})=p_\text{eq}+\frac{2\sigma_\text{eq}}{R}+\frac{\partial\sigma_\text{eq}}{\partial R} \,.
\label{eq_itsumrule3d}
\end{align}
It is easy to verify that, as in two dimensions, this sum rule is fulfilled for the active pressure, Eq.~\eqref{eq_pWofR}, together with the (negative) adsorption, Eq.~\eqref{eq_cavobsGAMMA3d}.
However, this is no longer the case for the active surface tension Eq.~\eqref{eq_cavobsSIGMA3d}, where $p^\text{(W-)}$ is only recovered up to the linear order in $R^{-1}$.
Also for a soft wall, the theoretical  adsorption and pressure are related via Eq.~\eqref{eq_itsumrule3d},
 coherent with the observations made in two dimensions.
This suggests that it is rather the adsorption than the active surface tension which within the EPA can be related to the active pressure by a sum rule, at least for an ideal gas and up to the leading terms in curvature.

\section{Normalization by total particle number \label{app_norm}}

In Sec.~\ref{sec_circular} we concluded that the chosen normalization of pressure and adsorption by a quantity only related to the bulk fraction of particles
leads to significant deviations for very persistent active particles.
Here we restate the results while normalizing by the total particle number $N=\int\upd \bvec{r} \rho(r)$.
For the adsorption, Eq.~\eqref{eq_gammaEFF}, this amounts to dividing by $N$ (and setting $A=|2\pi R|$), which yields
\begin{align}
\frac{\Gamma^{(-)}(R^{-1})}{N} &=\frac{1}{|2\pi R|}\,\frac{-\sqrt{\pi}\frac{l_\text{p}}{R}+\frac{l_\text{p}^2}{R^2}}{1-\sqrt{\pi}\frac{l_\text{p}}{R}+\frac{l_\text{p}^2}{R^2}}
\label{eq_cavobsGAMMAnorm}
\end{align}
 for a cavity. In the persistent limit, $l_\text{p}\rightarrow\infty$, we recover 
\begin{align}
 \Gamma^{(-)}(R^{-1})\rightarrow \frac{N}{|2\pi R|}\,, 
 \label{eq_GammaPERS}
\end{align}
which only depends on the curvature $|R^{-1}|$, in agreement with the prediction in Ref.~\onlinecite{fily2017}.

For the pressure, it appears sensible to normalize by $\Da N/V$, i.e., the active bulk pressure in a fluid with overall density $N/V\neq\rho_0$.
However, the result
\begin{align}
\frac{V \beta p^\text{(W-)}(R^{-1})}{\Da N}&=\frac{1-\frac{\sqrt{\pi}}{2}\frac{l_\text{p}}{R}}{1-\sqrt{\pi}\frac{l_\text{p}}{R}+\frac{l_\text{p}^2}{R^2}}\,,
\label{eq_cavobsnorm}
\end{align}
does not admit a finite value in the persistent limit as in Eq.~\eqref{eq_cavobsGAMMAnorm}.
This is because the volume $V=R^2\pi$ and the active diffusivity $\Da=l_\text{p}v_0\tau_0/(2d^2)$
depend on the cavity radius and the activity, respectively.
Therefore, we choose 
\begin{align}
\frac{\beta p^\text{(W-)}(R^{-1})}{N}&=\frac{v_0\tau_0}{d}\,\frac{1}{|2\pi R|d}\,\frac{-\frac{l_\text{p}}{R}+\frac{\sqrt{\pi}}{2}\frac{l_\text{p}^2}{R^2}}{1-\sqrt{\pi}\frac{l_\text{p}}{R}+\frac{l_\text{p}^2}{R^2}}\,,
\label{eq_cavobsnorm2}
\end{align}
which in the persistent limit,
\begin{align}
 \beta p^\text{(W-)}(R^{-1})\rightarrow 
 \frac{\sqrt{\pi}}{2}\,\frac{v_0\tau_0}{d}\,\frac{N}{|2\pi R|d}\,, 
 \label{eq_pPERS}
\end{align}
 depends linearly on both curvature and the (average) self-propulsion velocity.
The appearance of the latter quantity can be interpreted by some (average) momentum transferred to the wall by the adsorbed particles, cf., Eq.~\eqref{eq_GammaPERS}.

\begin{figure} [t] \centering
\includegraphics[width=0.45\textwidth] {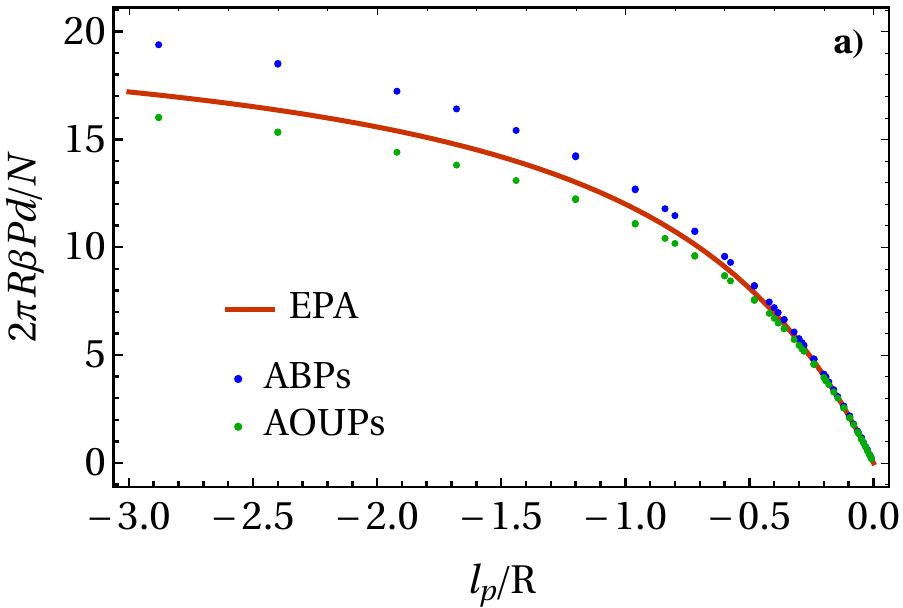}
\includegraphics[width=0.45\textwidth] {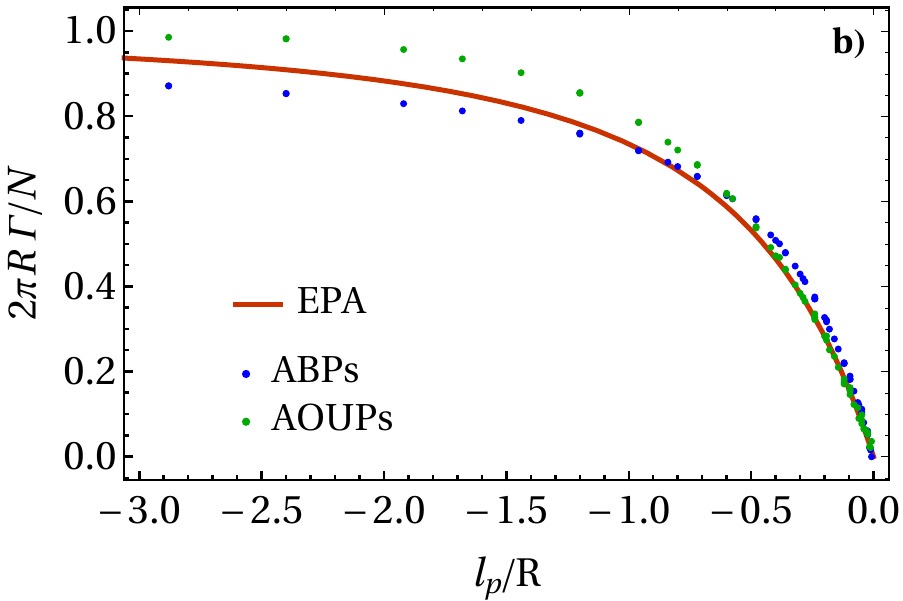}
\caption{
Renormalized mechanical properties of an active ideal gas on a (nearly) hard circular wall as a function of the ratio of persistence length $l_\text{p}$ and signed curvature radius $R$.
We display \textbf{(a)} the pressure $p(R^{-1})$ from Eq.~\eqref{eq_cavobsnorm2} with a self-propulsion velocity of $v_0=24d/\tau_0$
and \textbf{(b)} the adsorption $\Gamma(R^{-1})$ from Eq.~\eqref{eq_cavobsGAMMAnorm}.
For $R>0$ (obstacle) the values in both plots remain zero.
}\label{fig_idCapp}
\end{figure}

In Fig.~\ref{fig_idCapp} we plot the results of Eqs.~\eqref{eq_cavobsGAMMAnorm} and~\eqref{eq_cavobsnorm2} for a cavity and compare to the accordingly renormalized simulation data for ABPs and AOUPs. While for small curvatures, the behavior is reflected by the initial slope $m^p$, cf. table~\ref{tableSUM}, at larger curvatures we find slightly higher pressures and adsorptions for ABPs than for AOUPs, consistent with earlier work observing that AOUPs require lower degrees of activity in order to adsorb at the edge of a cavity \cite{fily2017}.
 For highly persistent particles, or at a strongly curved wall, all particles are trapped at the wall, such that $\Gamma$ satisfies Eq.~\eqref{eq_GammaPERS} in all models. Moreover, in this limit all particle velocities point directly towards the wall, resulting in a total integrated pressure $2\pi R \beta p = N \langle |v| \rangle \tau_0 /d^2$. Here, the mean absolute velocity $\langle v \rangle = v_0$ for ABPs, and $\langle v \rangle =v_0 \sqrt{\pi} /2$ for AOUPs. Hence, Eq.~\eqref{eq_pPERS} is satisfied exactly only for AOUPs, while ABPs exert a slightly higher pressure.

Applying the same normalizations for an obstacle would result in a theoretical adsorption and pressure, respectively, which are identically zero, 
since, in an unbounded system, there is an infinite number $N$ of particles, while both quantities are always finite within the normalization chosen in the main text.
However, the persistent limit for an obstacle is ill-defined in simulations, since boundaries are needed to contain the particles, but no length scale in the system should be smaller than the persistence length.
Nonetheless, since we know that no infinitely persistent particles can be found at a boundary with positive geometrical curvature, we can conclude that the theoretical prediction reflects the behavior in an active system and all curves in Fig.~\ref{fig_idCapp} can be trivially extended to $R>0$, with the flat-wall limit being not smooth at $R^{-1}$.
We thus conclude that, (only) upon the normalization by the total number of particles, both adsorption and pressure of an active ideal gas are captured exactly in the persistent limit for any wall potential, even if the local geometrical curvature is zero or positive.


\begin{thebibliography}{10}


\bibitem{palacci2013living} J.\ Palacci, S.\ Sacanna, A.\ P.\ Steinberg, D.\ J.\ Pine and P.\ M.\ Chaikin, Science \textbf{339}, 936 (2013).
\bibitem{buttinoni2013} I.\ Buttinoni, J.\ Bialk\'e, F.\ K\"ummel, H.\ L\"owen, C.\ Bechinger and T.\ Speck, Phys.\ Rev.\ Lett.\ \textbf{110}, 238301 (2013).

\bibitem{theurkauff2012dynamic} I.\ Theurkauff, C.\ Cottin-Bizonne, J.\ Palacci, C.\ Ybert and L.\ Bocquet, Phys.\ Rev.\ Lett.\ {\bf 108}, 268303 (2012).


\bibitem{cates_tailleur2014} M.~E.\ Cates and J.\ Tailleur, Annu.\ Rev.\ Condens.\ Matt.\ Phys.\ {\bf 6}, 219 (2015). 
\bibitem{takatori2015} S.\ C.\ Takatori and J.\ F.\ Brady, Phys.\ Rev.\ E \textbf{91}, 032117 (2015).

\bibitem{takatori2014} S.\ C.\ Takatori, W.\ Yan and J.\ F.\ Brady, Phys.\ Rev.\ Lett.\ \textbf{113}, 028103 (2014).
\bibitem{solon2015EOS} A.\ P.\ Solon, Y.\ Fily, A.\ Baskaran, M.\ E.\ Cates, Y.\ Kafri, M.\ Kardar and J.\ Tailleur, Nature Physics \textbf{11}, 673 (2015).
\bibitem{winkler2015} R.\ G.\ Winkler, A.\ Wysocki and G.\ Gompper, Soft Matter \textbf{11}, 6680 (2015).

\bibitem{speck_interface2014} J.~Bialk\'e, J.\ T.\ Siebert, H.\ L\"owen and T.\ Speck, Phys.\ Rev.\ Lett.\ \textbf{115}, 098301 (2015).
\bibitem{paliwal2017} S.\ Paliwal, V.\ Prymidis, L.\ Filion and M.\ Dijkstra, J.\ Chem.\ Phys.\ \textbf{147}, 084902 (2017).
\bibitem{Das2019} S.\ Das, G.\ Gompper and R.\ G.\ Winkler, arXiv preprint arXiv:1902.07435 (2019).

\bibitem{stenhammar2013} J.\ Stenhammar, A.\ Tiribocchi, R.\ J.\ Allen, D.\ Marenduzzo and M.\ E.\ Cates, Phys.\ Rev.\ Lett.\  {\bf 111}, 145702 (2013).
\bibitem{paliwal2018chemical} S.\ Paliwal, J.\ Rodenburg, R.\ van Roij and M.\ Dijkstra, New J.\ Phys.\ \textbf{20}, 015003 (2018).
\bibitem{meer2016chempot} B.\ van der Meer, V.\ Prymidis, M.\ Dijkstra and L.\ Filion, arXiv preprint arXiv:1609.03867 (2016).

\bibitem{loi2008effective} D.\ Loi, S.\ Mossa and L.\ F.\ Cugliandolo, Phys.\ Rev.\ E \textbf{77}, 051111 (2008).
\bibitem{szamel2014} G.\ Szamel, Phys.\ Rev.\ E, \textbf{90}, 012111 (2014).

\bibitem{wittmannbrader2016} R.\ Wittmann and J.~M.\ Brader, EPL \textbf{114}, 68004 (2016).
\bibitem{solon2018generalized} A.\ P.\ Solon, J.\ Stenhammar, M.\ E.\ Cates, Y.\ Kafri and J.\ Tailleur, Phys.\ Rev.\ E, \textbf{97}, 020602 (2018).

\bibitem{solon_BrownianPressure2015} A.\ P.\ Solon, J.\ Stenhammar, R.\ Wittkowski, M.\ Kardar, Y.\ Kafri, M.\ E.\ Cates and J.\ Tailleur, Phys.\ Rev.\ Lett.\ \textbf{114}, 198301 (2015).

\bibitem{smallenburg2015} F.\ Smallenburg and H.\ L\"owen, Phys.\ Rev.\ E \textbf{92}, 032304 (2015).
\bibitem{nikola2016} N.\ Nikola, A.\ P.\ Solon, Y.\ Kafri, M.\ Kardar, J.\ Tailleur and R.\ Voituriez, Phys.\ Rev.\ Lett.\ \textbf{117}, 098001 (2016).

\bibitem{faragebrader2015} T.~F.~F.\ Farage, P.~Krinninger and J.~M.\ Brader, Phys.\ Rev.\ E \textbf{91}, 042310 (2015). 
\bibitem{fodor2016} E.\ Fodor, C.\ Nardini, M.\ E.\ Cates, J.\ Tailleur, P.\ Visco and F.\ van Wijland, Phys.\ Rev.\ Lett.\ \textbf{117} 038103 (2016).

\bibitem{solonEPJST} A.\ P.\ Solon, M.\ E.\ Cates and J.\ Tailleur, Eur.\ Phys.\ J.\ Special Topics \textbf{224}, 1231 (2015).


\bibitem{fily2017} Y.\ Fily, A.\ Baskaran and M.\ F.\ Hagan, Eur.\ Phys.\ J.\ E \textbf{40}, 61 (2017).

\bibitem{Das2018} S.\ Das, G.\ Gompper and R.\ G.\ Winkler, New J.\ Phys.\ \textbf{20}, 015001 (2018). 

\bibitem{sandford2017} C.\ Sandford, A.\ Y.\ Grosberg and J.-F.\ Joanny, Phys.\ Rev.\ E \textbf{96}, 052605 (2017).
\bibitem{sandford2018} C.\ Sandford and A.\ Y.\ Grosberg, Phys.\ Rev.\ E \textbf{97}, 012602 (2018).



\bibitem{maggi2015sr} C.~Maggi, U.~M.~B.\ Marconi, N.\ Gnan and R.\ Di Leonardo, Scientific Reports \textbf{5} 10742 (2015). 
\bibitem{marconi2015} U.~M.~B.\ Marconi and C.~Maggi, Soft Matter \textbf{11}, 8768 (2015). 
\bibitem{marconi2016} U.~M.~B.\ Marconi, C.~Maggi and S.\ Melchionna, Soft Matter \textbf{12}, 5727 (2016).

\bibitem{SpeckCRIT} M.\ Rein and T.\ Speck, Eur.\ Phys.\ J.\ E \textbf{39}, 84 (2016).

\bibitem{sharma2017} A.\ Sharma, R.\ Wittmann and J.~M.\ Brader, Phys.\ Rev.\ E, \textbf{95}, 012115 (2017).

\bibitem{marconi2016mp} U.\ M.\ B.\ Marconi, M.\ Paoluzzi and C.\ Maggi, Mol.\ Phys.\ \textbf{114}, 2400 (2016).

\bibitem{activePAIR} R.\ Wittmann, C.\ Maggi, A.\ Sharma, A.\ Scacchi, J.\ M.\ Brader and U.\ M.\ B.\ Marconi, J.\ Stat.\ Mech.\ \textbf{2017}, 113207.
\bibitem{activePRESSURE} R.\ Wittmann, U.\ M.\ B.\ Marconi, C.\ Maggi and J.\ M.\ Brader, J.\ Stat.\ Mech.\ \textbf{2017}, 113208.

\bibitem{activeMixture}  R.\ Wittmann, J.\ M.\ Brader, A.\ Sharma and U.\ M.\ B.\ Marconi, Phys.\ Rev.\ E \textbf{97}, 012601 (2018).
\bibitem{marconiExactpressure2017} U.~M.~B.\ Marconi, C.\ Maggi and M.\ Paoluzzi, J.\ Chem.\ Phys.\ \textbf{147}, 024903 (2017).

\bibitem{paoluzzi2018} M.\ Paoluzzi, U.\ M.\ B.\ Marconi and C.\ Maggi, Phys.\ Rev.\ E \textbf{97}, 022605 (2018).
\bibitem{caprini2018} L.\ Caprini and U.\ M.\ B.\ Marconi, Soft Matter \textbf{14}, 9044 (2018).
\bibitem{fily2018} Y.\ Fily, arXiv preprint arXiv:1812.05698 (2018).

\bibitem{archerevans2017} A.\ J.\ Archer, B.\ Chacko and R.\ Evans, J.\ Chem.\ Phys.\ \textbf{147}, 034501 (2017).

\bibitem{faetti1988} S.\ Faetti, L.\ Fronzoni, P.\ Grigolini and R.\ Mannella, Journal of Statistical Physics \textbf{52}, 951 (1988).
\bibitem{UCNA} P.\ H\"anggi and P.\ Jung, Adv.\ Chem.\ Phys.\ \textbf{89}, 239 (1995).

\bibitem{clarifyST} Although $\sigma$, as defined in Eq.~\eqref{eq_gamma}, is usually referred to as surface tension, it differs from the actual surface tension, defined as an integral of the pressure tensor anisotropy, by an additional wall term. 
This can be understood by rewriting the integral over $p\,\Theta(\mp(r-R))$ in terms of the nomal pressure,
which vanishes inside the wall and equals $p$ in the bulk, and using the condition $\nab\cdot\boldsymbol{p}=-\rho\nab \Vext$ for hydrostatic stability.
The additional wall term does, however, only play a role if the wall potential is soft,
which means that at a hard wall $\sigma$ equals the surface tension.
Due to this minor difference, we shortly call $\sigma$ a \textit{total} surface tension,
or just surface tension if there is no risk of confusion with the true surface tension.
%
For a passive system, the true surface tension is a pure many-body term, which is always zero for an ideal gas, whereas the total surface tension, i.e., the excess grand potential per surface area $\sigma$ determines surface thermodynamics, compare Ref.~\onlinecite{sumrules}.
\bibitem{sumrules} J.\ R.\ Henderson, Ch.2 in {\it Fundamentals of inhomogeneous fluids} (Marcel Dekker) 1992. 

\bibitem{FOX} R.\ F.\ Fox, Phys.\ Rev.\ A \textbf{33}, 467 (1986).



\bibitem{varnik2000molecular} F.\ Varnik, J.\ Baschnagel and K.\ Binder, J.\ Chem.\ Phys.\ \textbf{113}, 4444 (2000).

\bibitem{kirkwoodbuff} J.\ G.\ Kirkwood and F.\ B.\ Buff, J.\ Chem.\ Phys.\ \textbf{17}, 338 (1949).


\end{thebibliography}
\end{document}